\documentclass{aa}  

\usepackage{graphicx}
\usepackage{xcolor}
\usepackage{txfonts}
\renewcommand{\textbf}[1]{#1}

\newcommand{\g}{{\rm g}}

\newcommand{\cm}{{\rm cm}}

\newcommand{\au}{{\rm ~au}}

\newcommand{\yr}{{\rm yr}}

\begin{document}

    \title{Fate of a remnant solid disk around an eccentric giant planet} 


   \author{S. Shibata
          \inst{1,2}
          \and
          R. Helled\inst{1}
          }

   \institute{Department for Astrophysics, University of Zurich, Switzerland
              \and
              Department of Earth, Environmental and Planetary Sciences, Rice University, USA
              \\
              \email{s.shibata423@gmail.com}              
             }

   \date{}

 
  \abstract
   {
    The composition of giant planets'  atmospheres is an important tracer of their formation history.  
    While many theoretical studies investigate the heavy-element accretion within a gaseous protoplanetary disk, the possibility of solid accretion after disk dissipation has not been explored.  
    }
   {
    Here, we focus on the case of a gas giant planet excited to an eccentric orbit and assess the likelihood of solid accretion after disk dissipation. We follow the orbital evolution of the surrounding solid materials and investigate the scattering and accretion of heavy elements in the remnant solid disks.
    }
   {
    We perform N-body simulations of planetesimals and embryos around an eccentric giant planet.
    We consider various sizes and orbits for the eccentric planet and determine the fate of planetesimals and embryos. 
    }
   {
    We find that the orbital evolution of solids, such as planetesimals and embryos, is regulated by weak encounters with the eccentric planet rather than strong close encounters.
    Even in the region where the Safronov number is smaller than unity, most solid materials fall onto the central star or are ejected from the planetary system. 
    We also develop an analytical model of the solid accretion along the orbital evolution of a giant planet, where the accretion probability is obtained as a function of the planetary mass, radius, semi-major axis, eccentricity, inclination, and solid disk thickness.
   }
   {
    Our model predicts that $\sim$0.01-0.1 $M_\oplus$ of solids is accreted onto an eccentric planet orbiting in the outer disk ($\sim10\au$). 
    The accreted heavy-element mass increases (decreases) with the eccentricity (inclination) of the planet. 
    We also discuss the possibility of collisions of terrestrial planets and find that $\sim10\%$ of the hot Jupiters formed via high-eccentric migration collide with a planet of $10M_\oplus$.
    However, we find that solid accretion and collisions with terrestrial planets are minor events for planets in the inner orbit, and a different accretion process is required to enrich eccentric giant planets with heavy elements.      
   }


   \maketitle
%

\section{Introduction}
The formation history of giant planets is still not fully understood. 
Many theoretical studies investigate the accretion of heavy elements onto the planet at various stages since the observed atmospheric composition of giant planets is thought to be related to the formation history.
Heavy-element accretion is now known to occur mainly before the dissipation of the protoplanetary disk. 
The accretion of dust \citep{Morbidelli+2023}, planetesimals \citep{Shibata+2020, Shibata+2022a, Hands+2021, Turrini+2021, Knierim+2022, Pacetti+2022}, and disk gas accretion \citep{Madhusudhan+2014, Booth+2017, Booth+2019, Notsu+2020, Schneider+2021, Schneider+2021b} are all regarded to be critical processes that can affect the planetary composition. 
The above studies suggest that the ratio of molecular species in the planetary atmosphere could be a tracer of the initial formation location of giant planets.
However, these studies focus on the accretion of heavy elements before the disk dissipates. 
In reality, accretion could also occur at later stages and change the composition of the planetary atmosphere. 
Therefore, it is important to investigate the effects of solid accretion after disk dissipation to link the initial formation location of giant planets to their atmospheric composition.

Heavy-element accretion post-disk dissipation depends on the orbital evolution of giant planets and the remnant solid disk. 
By the end of the disk dissipation, solid particles like planetesimals are eliminated from a feeding zone due to the aerodynamical drag from the protoplanetary disk gas.  
This means that no solid accretion occurs without the scattering process triggered by the orbital instability of planets.
Planetesimal scattering by giant planets is typically investigated using N-body simulations \citep{Matter+2009, Frewen+2014, Mustill+2018, Seligman+2022, Rodet+2023}.
\citet{Seligman+2022} consider the accretion of comets, which are planetesimals scattered by the other giant planets.
Their analytical model suggests that a close-in giant planet can accrete a significant fraction of comets before they are scattered from the planetary system. 
However, they assume two groups of giant planets: (i) 
giant planets that scatter the remnant solid disk in the outer disk region,
and (ii) giant planets that accrete the scattered solid disk in close-in orbits. 
Before being accreted by the inner giant planets, some solids can be accreted by the outer giant planets. 

\citet{Rodet+2023} investigate the scattering and accretion of planetesimals around an eccentric giant planet. 
They find that the planet can accrete some scattered planetesimals and that the accretion probability changes with the planet's semi-major axis, eccentricity, mass, and radius.
However, this study focuses on the planetesimals that fall onto the central star.
The initial inclination of the planetesimal disk and the mutual inclination between the planetesimal disk and the planet should affect the accretion probability, but this has not been investigated in previous studies. 
Also, they start the simulations with the planet whose orbit was already crossed to the planetesimal disk.
Since the planetary eccentricity is gradually excited by the mutual interactions with the other planets after the disk dissipation \citep{Carrera+2019, Anderson+2019, Garzon+2022}, in reality, the scattering of the remnant solid disk starts before the orbits are crossed.
It is, therefore, still unknown how much solid material can be accreted by the planet during the scattering process. 

In order to investigate the possible accretion of a remnant solid disk, we perform N-body simulations and analyze how the orbits of planetesimals and embryos evolve around an eccentric giant planet.
\textbf{
First, we derive the analytical equation for the accretion probability of solid particles onto an eccentric giant planet in Sec.~\ref{sec: CollisionToPlanet_ANL}.
In Sec.~\ref{sec: CollisionToPlanet_SIM}, we compare the derived formulae to the N-body simulations.}
Following \citet{Rodet+2023}, we begin the simulations with a planet with a finite eccentricity.
\textbf{
In order to model the evolution of giant planets in more detail, we investigate the accretion probability when the planetary eccentricity increases with time in Sec.~\ref{sec: EccentricityEvolution}.
}
We discuss our results, the limitations of our analytical model,  and the effects of solid accretion on the planetary composition in section~\ref{sec: Discussion}. 
We summarize our results and present the key conclusions in section~\ref{sec: Summary}.

\section{Accretion probability to an eccentric planet: Analytical investigations}
\label{sec: CollisionToPlanet_ANL}
\textbf{
We start with a simple situation where a giant planet crosses the orbits of surrounding planetesimals, following \citep{Rodet+2023}.
We neglect the mutual interaction between planetesimals; therefore, our analysis can be applied to the smaller solid particles, such as debris and dust.
The mass $M_\mathrm{p}$, radius $R_\mathrm{p}$, semi-major axis $a_\mathrm{p}$, eccentricity $e_\mathrm{p}$, and inclination $i_\mathrm{p}$ of the planet are fixed values.
}

\subsection{Orbital evolution around an eccentric giant planet}\label{sec:standard}
First, we consider the gravitational perturbation from the eccentric planet to the planetesimals
\textbf{
at the orbital radius $r$.
}
In a-particle-in-a-box approximation, a velocity perturbation by the passing planet is given by \citep[e.g.][]{Armitage2010}: 
\begin{eqnarray}
    \delta v \sim \frac{2 \mathcal{G} M_\mathrm{p}}{b v_\mathrm{rel}}, \label{eq:deltav}
\end{eqnarray}    
where $\mathcal{G}$ is the gravitational constant, $b$ is the impact parameter, and $v_\mathrm{rel}$ is the relative velocity to the passing planet.
Here, we assume that the impact parameter is the order of $r$, and the relative velocity is $e_\mathrm{p} v_\mathrm{K} (r)$, 
\textbf{
where $v_\mathrm{K}$ is the Kepler velocity.
}
The weak perturbation from the passing planet is given by: 
\begin{eqnarray}
    \delta v_\mathrm{weak} \sim v_\mathrm{K} (r) \frac{2}{e_\mathrm{p}}  \frac{M_\mathrm{p}}{M_\mathrm{s}}, \label{eq:deltav_far}
\end{eqnarray}
and the eccentricity perturbation on a planetesimal is approximated to be:  
\begin{eqnarray}
    \delta e_\mathrm{weak} \sim C_\mathrm{v} \frac{\delta v_\mathrm{weak}}{v_\mathrm{K} (r)} = \frac{2 C_\mathrm{v}}{e_\mathrm{p}} \frac{M_\mathrm{p}}{M_\mathrm{s}}, \label{eq:decc_weak}
\end{eqnarray}
where $C_\mathrm{v}$ is a fitting parameter.
The eccentricity excitation occurs in every orbit of the eccentric planet, and the excitation rate is independent of the radial distance from the central star.
Due to the large mass difference between the planet and planetesimals, the perturbations increase the mean eccentricity of the planetesimals $\left\langle e \right\rangle$ linearly.
The eccentricity evolution can be written as:
\begin{eqnarray}
    \frac{\mathrm{d} \left \langle e \right \rangle}{\mathrm{d} t} &\sim& \frac{\Delta e}{\Delta t} = \frac{\delta e_\mathrm{weak}}{P_\mathrm{orb,p}}, \\
    \left \langle e \right \rangle &\sim& \left \langle {e_0} \right \rangle + \delta e_\mathrm{weak} \frac{t}{P_\mathrm{orb,p}}, \label{eq:TimeEvolution_Ecc}
\end{eqnarray}
where $P_\mathrm{orb,p}$ is the orbital period of the eccentric planet
\textbf{
and $\left \langle {e_0} \right \rangle$ is the initial mean eccentricity of planetesimals.
}

Different from eccentricity excitation, inclination excitation requires close encounters with the eccentric planet \citep[e.g.][]{Goldreich+2004}.
When the eccentric planet passes through the planetesimal disk with $\sin i_\mathrm{p}<\sin i$, a z-component of the velocity perturbation is given by: 
\begin{eqnarray}
    \delta v_\perp = \delta v \frac{r \sin i}{b}.
\end{eqnarray}
We define the distance of the close encounter $b_\mathrm{close}$ as the distance at which the velocity perturbation doubles the planetesimal's inclination.
Therefore, 
\begin{eqnarray}
    C_\mathrm{v} \frac{\delta v_\perp}{v_\mathrm{K}} &=& \sin i, \\
    \frac{b_\mathrm{close}}{r} &=& \sqrt{\frac{2 C_\mathrm{v}}{e_\mathrm{p}} \frac{M_\mathrm{p}}{M_\mathrm{s}}}, \\
                     &=& 0.062 \left(\frac{C_\mathrm{v}}{2}\right)^{1/2} \left(\frac{e_\mathrm{p}}{0.99} \right)^{-1/2} \left( \frac{M_\mathrm{p}}{M_\mathrm{Jup}} \right)^{1/2} \left( \frac{M_\mathrm{s}}{M_\odot} \right)^{-1/2}.
\end{eqnarray}
At the beginning of the simulation, the thickness of the planetesimal disk $a \sin i$ is smaller than $b_\mathrm{close}$.
The typical timescale of the close encounter $\tau_\mathrm{close,2D}$ is given by: 
\begin{eqnarray}
    \tau_\mathrm{close,2D} &=& \frac{2 \pi r}{2 b_\mathrm{close}} P_\mathrm{orb,p}, \\
    &\sim& 51 P_\mathrm{orb,p} \left(\frac{C_\mathrm{v}}{2}\right)^{-1/2} \left(\frac{e_\mathrm{p}}{0.99} \right)^{1/2} \left( \frac{M_\mathrm{p}}{M_\mathrm{Jup}} \right)^{-1/2} \left( \frac{M_\mathrm{s}}{M_\odot} \right)^{1/2}. \label{eq:tau_close_2D}
\end{eqnarray}
Close encounters occur in three dimensions once the disk thickness exceeds $b_\mathrm{close}$.  
In this case, the typical timescale of the close encounter $\tau_\mathrm{close,3D}$ is given by: 
\begin{eqnarray}
    \tau_\mathrm{close,3D} = \frac{2 \pi r \cdot r \sin i}{\pi {b_\mathrm{close}}^2 } P_\mathrm{orb,p}. \label{eq:tau_close_3D}
\end{eqnarray}
The time evolution of mean inclination $\left \langle \sin i \right \rangle$ can be written as follows: 
\begin{eqnarray}
    \frac{ \mathrm{d} \left \langle \sin i \right \rangle}{\mathrm{d} t} \sim \frac{\Delta \sin i}{\Delta t} = 
    \left\{  \begin{array}{ll}
        \frac{\left \langle \sin i \right \rangle}{\tau_\mathrm{close,2D}} & \mathrm{ for}~ \left \langle \sin i \right \rangle < \sin i_\mathrm{b} \\
        \frac{\left \langle \sin i \right \rangle}{\tau_\mathrm{close,3D}} & \mathrm{ for}~ \sin i_\mathrm{b} < \left \langle \sin i \right \rangle 
    \end{array} \right. , \\
    \left \langle \sin i \right \rangle \sim
        \left\{  \begin{array}{ll}
            \left \langle \sin i_0 \right \rangle \exp \left( \frac{t}{\tau_\mathrm{close,2D}} \right) & \mathrm{ for}~ \left \langle \sin i \right \rangle < \sin i_\mathrm{b} \\
            \frac{ {b_\mathrm{close}}^2}{4 r^2} \frac{t-t_\mathrm{b}}{P_\mathrm{orb,p}} + \sin i_\mathrm{b} & \mathrm{ for}~ \sin i_\mathrm{b} < \left \langle \sin i \right \rangle 
            \end{array} \right. ,
    \label{eq:TimeEvolution_Inc}
\end{eqnarray}    
with
\begin{eqnarray}
    \sin i_\mathrm{b} &=& \frac{b_\mathrm{close}}{r}, \\
    t_\mathrm{b} &=& \tau_\mathrm{close, 2D} \log \frac{\sin i_\mathrm{b}}{\left \langle \sin i_0 \right \rangle}.
\end{eqnarray}
\textbf{
Here, $\left \langle \sin i_0 \right \rangle$ is the mean initial inclination of planetesimals.
}

\subsection{Accretion to close encounter ratio}\label{sec:fcc}
The planetesimal accretion probability depends on the geometry of the planetesimal disk.
The capture radius of an eccentric planet is given by: 
\begin{eqnarray}
    R_\mathrm{cap} = R_\mathrm{p} F_\mathrm{grav}  
                   = R_\mathrm{cap} \left( 1 + \frac{ {v_\mathrm{esc}}^2 }{ 2 {v_\mathrm{rel}}^2 } \right)^{1/2},
\end{eqnarray}
where $F_\mathrm{grav}$ is the gravitational focusing factor, $v_\mathrm{esc}$ and $v_\mathrm{rel}$ are the escape velocity of the planet and the relative velocity between the planet and planetesimals.
When the half thickness of the planetesimal disk $r \sin i$ is smaller than the capture radius $R_\mathrm{cap}$, particle accretion occurs two-dimensionally (2D).
The accretion probability per orbit of the eccentric planet  is given by: 
\begin{eqnarray}
    f_\mathrm{acc,2D} = \frac{R_\mathrm{cap}}{\pi r} \label{eq:fcol_2D}.
\end{eqnarray}
Once the thickness of the particle disk exceeds  $R_\mathrm{cap}$, the particle collision occurs in a 3D manner. 
The accretion probability is then given by: 
\begin{eqnarray}
    f_\mathrm{acc,3D} =  \frac{{R_\mathrm{cap}}^2}{4 \pi r^2 \sin i} \label{eq:fcol_3D}.
\end{eqnarray}

The close encounters with the eccentric planet increase the planetesimals' inclination.
The close encounter probability per planetary orbit is given by: 
\begin{eqnarray}
    f_\mathrm{close,2D} = \frac{b_\mathrm{close}}{\pi r}. \label{eq:fclose_2D}
\end{eqnarray}
When the thickness of the planetesimal disk exceeds $b_\mathrm{close}$, the close encounter probability is:
\begin{eqnarray}
    f_\mathrm{close,3D} =  \frac{{b_\mathrm{close}}^2}{4 \pi r^2 \sin i}. \label{eq:fclose_3D}
\end{eqnarray}

Using Eqs~(\ref{eq:fcol_2D}), (\ref{eq:fcol_3D}), (\ref{eq:fclose_2D}), and (\ref{eq:fclose_3D}), we obtain the accretion fraction per single close encounter $f_\mathrm{a/c}$ to be: 
\begin{eqnarray}\label{eq:collision_to_scattering}
    f_\mathrm{a/c} =
        \begin{cases}
            \frac{f_\mathrm{acc,2D}}{f_\mathrm{close,2D}} = \frac{R_\mathrm{cap}}{b_\mathrm{close}} &\text{~for~} \sin i < \frac{R_\mathrm{cap}}{r}, \\
            \frac{f_\mathrm{acc,3D}}{f_\mathrm{close,2D}} = \frac{R_\mathrm{cap}}{b_\mathrm{close}} \frac{R_\mathrm{cap}}{r \sin i} &\text{~for~} \frac{R_\mathrm{cap}}{r} < \sin i < \frac{b_\mathrm{close}}{r}, \\
            \frac{f_\mathrm{acc,3D}}{f_\mathrm{close,3D}} = \left( \frac{ R_\mathrm{cap} }{b_\mathrm{close}} \right)^2 &\text{~for~} \frac{b_\mathrm{close}}{r} < \sin i. \\       
        \end{cases}
\end{eqnarray}

\subsection{Accretion probability}
The accretion probability of planetesimals onto an eccentric planet $P_\mathrm{col}^\prime$ is given by:
\begin{eqnarray}\label{eq:Pcol}
    P^\prime_\mathrm{col} &= \min( 1, P^{\prime\prime}_\mathrm{col}),
\end{eqnarray}
with
\begin{eqnarray}
    P^{\prime\prime}_\mathrm{col} &=& \int_0^{\tau_\mathrm{eject}} \frac{f_{a/c}}{\tau_\mathrm{close}} \mathrm{d} t, \label{eq: Pcol_int}
\end{eqnarray}
where $\tau_\mathrm{eject}$ is the mean duration time of the planetesimals before they are eliminated.
If the eccentricity exceeds $e_\mathrm{fall}=1-R_\mathrm{star}/a$, 
\textbf{
where $R_\mathrm{star}$ is the radius of the central star and $a$ is the semi-major axis of the planetesimals,
}
the planetesimal will be removed from the planetary system by falling to the central star or ejection from the planetary system.
By equating the mean eccentricity given by eq.~(\ref{eq:TimeEvolution_Ecc}) to $e_\mathrm{fall}$, we obtain $\tau_\mathrm{eject}$ as follows: 
\begin{eqnarray}
    \frac{\tau_\mathrm{eject}}{P_\mathrm{orb,p}} &=& \frac{e_\mathrm{p}}{2 C_\mathrm{v}} \frac{M_\mathrm{s}}{M_\mathrm{p}} \left\{ \left( 1 - \frac{R_\mathrm{star}}{a} \right) - \left\langle e_0 \right \rangle \right\}. \label{eq:stay_time}
\end{eqnarray}
\textbf{
Using the initial semi-major axis of a planetesimal $a_0$, we integrate eq.~(\ref{eq: Pcol_int}) and obtain
}
\begin{eqnarray}
    P^{\prime\prime}_\mathrm{col} = P_\mathrm{2D} +P_\mathrm{3D2D} +P_\mathrm{3D}, \label{eq: Pcol_pp}
\end{eqnarray}
with
\begin{align}
    P_\mathrm{2D} =
    \begin{cases}
          \frac{R_\mathrm{cap}}{b_\mathrm{close}} \log \frac{R_\mathrm{cap}}{a_0 \left \langle \sin i_0 \right \rangle}
        &\text{~for~} \left \langle \sin i_0 \right \rangle < \frac{R_\mathrm{cap}}{a_0}, \\
         0
        &\text{~other~}. \\
    \end{cases} 
\end{align}
\begin{align}
    P_\mathrm{3D2D} =
    \begin{cases}
          \frac{R_\mathrm{cap}}{b_\mathrm{close}} \left(1-\frac{R_\mathrm{cap}}{b_\mathrm{close}} \right)
        &\text{~for~} \sin i_0 < \frac{R_\mathrm{cap}}{a_0}, \\
          \frac{R_\mathrm{cap}}{b_\mathrm{close}} \left(\frac{R_\mathrm{cap}}{a_0 \sin i_0 }-\frac{R_\mathrm{cap}}{b_\mathrm{close}} \right)
        &\text{~for~} \frac{R_\mathrm{cap}}{a_0} < \sin i_0 < \frac{b_\mathrm{close}}{a_0}, \\
         0
        &\text{~other~}. \\
    \end{cases} 
\end{align}
\begin{align}
    P_\mathrm{3D} =
    \begin{cases}
          \left( \frac{R_\mathrm{cap}}{b_\mathrm{close}} \right)^2 
            &\log \left\{ \frac{1}{4} \frac{b_\mathrm{close}}{a_0} \left( \frac{\tau_\mathrm{eject}}{P_\mathrm{orb,p}}-\pi \log \frac{\sin i_\mathrm{b}}{ \sin i_0 } \right) + 1 \right\} \\
          &\text{~for~} \sin i_0 < \frac{b_\mathrm{close}}{a_0}, \\
          \left( \frac{R_\mathrm{cap}}{b_\mathrm{close}} \right)^2 
            &\log \left\{ \frac{1}{4 \sin i_0 } \left(\frac{b_\mathrm{close}}{a_0}\right)^2 \frac{\tau_\mathrm{eject}}{P_\mathrm{orb,p}} + 1 \right\} \\
          &\text{~other~}. \\
    \end{cases}
\end{align}
\textbf{
We assume that the orbital distance of planetesimals $r$ does not change from $a_0$ during the scattering by the eccentric planet.
This is not true for each planetesimal, but numerical simulations in Sec.~\ref{sec: CollisionToPlanet_SIM} show that this could be a reasonable assumption by taking an average for a large number of planetesimals.
}

\textbf{
The inferred accretion probability depends on $R_\mathrm{cap}/b_\mathrm{close}$, that is
\begin{align}    
    R_\mathrm{cap}/b_\mathrm{close} &= \frac{R_\mathrm{p}}{r} \sqrt{ \frac{e_\mathrm{p}}{2 C_\mathrm{v}} \frac{M_\mathrm{s}}{M_\mathrm{p}} } \left( 1 +\frac{\Theta}{{e_\mathrm{p}}^2} \right)^{1/2}, \\
    &\propto
    \begin{cases}
        {e_\mathrm{p}}^{-1/2} \left( \frac{R_\mathrm{p}}{r} \right)^{1/2} &\text{~for~} e_\mathrm{p} \ll \sqrt{\Theta}, \\
        {e_\mathrm{p}}^{1/2} \left( \frac{R_\mathrm{p}}{r} \right) \left( \frac{M_\mathrm{p}}{M_\mathrm{s}} \right)^{-1/2} &\text{~for~} e_\mathrm{p} \gg \sqrt{\Theta},
    \end{cases}
\end{align}
where $\Theta$ is the Safronov number and given as
\begin{align}
    \Theta = v_\mathrm{esc}^2 / 2 v_\mathrm{K}^2.
\end{align}
}
The Safronov number is usually used as an indicator for describing the fate of planetary objects in an unstable system.
An encounter grazing the planetary surface puts a velocity kick of $\sim v_\mathrm{esc}$ to a planetesimal, which leads to $\delta e \sim \sqrt{2 \Theta}$. 
If the Safronov number is larger than one, the gravitational scattering by the planet is likely to lead to ejection. 
Therefore, it is predicted that accretion is dominant in the region of $\Theta<1$ while ejection or falling into the star is dominant with $\Theta>1$.
\textbf{
Our model suggests that the accretion probability is affected by not only the Safronov number but also the planet's mass, radius, eccentricity, and planetesimals' initial inclination.
}



\section{Accretion probability to an eccentric planet: Numerical simulations}
\label{sec: CollisionToPlanet_SIM}
In the previous section, we derived an analytical formula for the accretion probability of planetesimals. 
We found that the accretion probablity depends on not only $M_\mathrm{p}, R_\mathrm{p}, e_\mathrm{p}$ as found in previous studies, but also $i_\mathrm{0}$.
To assess the obtained accretion probability formula, we investigate the orbital evolution of planetesimals using N-body simulations.

\subsection{Numerical Setup}
We integrate orbits of planetesimals around a central star with a mass of $M_\mathrm{s}$ and with an eccentric protoplanet with a mass of $M_\mathrm{p}$. 
The planet's semi-major axis $a_\mathrm{p}$ and the eccentricity $e_\mathrm{p}$ are set as input parameters.
In this case, the perihelion and aphelion of the planet are $a_\mathrm{p} (1-e_\mathrm{p})$ and $a_\mathrm{p} (1+e_\mathrm{p})$.
The inclination $\sin i_\mathrm{p}$ is also an input parameter.
The other orbital angles, longitude of ascending node $\Omega_\mathrm{p}$, longitude of pericentre $\varpi_\mathrm{p}$, and true anomaly $\tau_\mathrm{p}$, are set as $(\Omega_\mathrm{p},\varpi_\mathrm{p},\tau_\mathrm{p})=(0,0,\pi)$ or $(0,1/2\pi,\pi)$.
We distribute planetesimals around the eccentric planet and integrate the orbits up to a time  $t=10^6 \yr$, or until all the particles other than the planet have been eliminated from the simulations.
In our simulations, we use the IAS 15 integrator \citep{Everhart+1985, Rein+2015}, whose accuracy is sufficiently high to calculate the orbital evolution of eccentric objects. 
The benchmark results of our simulations are described in appendix \ref{app: Integrator}. 

We count the number of particles that collide with the eccentric protoplanet $N_\mathrm{col}$, fall to the central star $N_\mathrm{fall}$, and are ejected from the planetary system $N_\mathrm{eject}$.
We judge that a particle collides with the eccentric protoplanet once the relative distance becomes smaller than the physical radius of the planet $R_\mathrm{p}$.
A particle whose orbital distance from the central star $r$ becomes smaller than the central star's radius $R_\mathrm{star} = 0.0046 \au$ (comparable to the solar radius) is assumed to fall into the central star. 
Once $r>10^3 \au$ is achieved, the particle is assumed to be eliminated from the planetary system.


We assume that the planetesimals are distributed uniformly in logarithmic space in the radial direction. 
We distribute planetesimals from $a_\mathrm{pl,in}$ to $a_\mathrm{pl,out}$, where $a_\mathrm{pl,in}$ is set as smaller than $a_\mathrm{p}(1-e_\mathrm{p})$, and $a_\mathrm{pl,out}$ is set as larger than $a_\mathrm{p}(1+e_\mathrm{p})$.
The eccentricity and inclination are then given by Rayleigh distribution with the eccentricity dispersion $\langle e_0^2 \rangle^{1/2}$ and inclination dispersion $\langle \sin ^2 i_0 \rangle^{1/2}$.
The other orbital angles of planetesimals, the longitude of ascending node $\Omega$, the longitude of pericentre $\varpi$, and true anomaly $\tau$, are chosen randomly.

To reduce the computational cost, the planetesimals are treated as test particles. 
Since planetesimals are represented by test particles, our results can also be applied to solid disks of smaller particles like debris and dust.
The number of test particles $N_\mathrm{tp}$ is set so that more than 400 particles are in each bin with a width of 0.1 in logarithmic space in the radial direction. 
\textbf{
The surface density of the planetesimals goes as $\propto r^{-2}$.
Note that the density profile  of the solid disk is not necessarily the same as our planetesimal disk. 
The  goal of our numerical simulations is to infer the accretion probability, which does not depend on the solid disk profile as long as the mutual interactions are negligible.
In this studuy, we assume that the planetesimal disk has the same number density in each bin. In any case, the accretion probablity we derive is independent of the assumed disk structure. However, the total accreted heavy-element mass would change for different assumed disk profiles and could be investigated in future research.}
At the end of simulations, some planetesimals remain in the system.
However, we stop the simulations at $t=10^6 \yr$ because planetesimal accretions with the protoplanet mainly occur around $t\lesssim10^5~\yr$. 
Therefore, our numerical results are unexpected to change if longer simulation times are considered. 

\begin{table*}
	\centering
	\begin{tabular}{ c|c|c|c } 
		\hline
		& parameter & standard value & parameter study \\
		\hline
        $M_\mathrm{s}$   & mass of central star & $M_\odot$ & \\
    	$M_\mathrm{p}$   & mass of protoplanet & $M_\mathrm{Jup}$ & $1/8 M_\mathrm{Jup}$-$8 M_\mathrm{Jup}$  \\
    	$R_\mathrm{p}$   & radius of protoplanet & $R_\mathrm{Jup}$ & $1/8 R_\mathrm{Jup}$-$R_\mathrm{Jup}$ \\
    	$a_\mathrm{p}$   & semi-major axis of protoplanet & $1 \au$ & $1 \au$, $10 \au$ \\
    	$e_\mathrm{p}$   & eccentricity of protoplanet & $0.99$ & $0.1$-$0.99$ \\
    	$\sin i_\mathrm{p}$   & inclination of protoplanet & $0$ & $0,10^{-3},10^{-2}$ \\
    	$\varpi_\mathrm{p}$ & longitude of pericenter of protoplanet & $0$ & $0,1/2\pi$ \\
    		\hline
		$a_\mathrm{pl,in}$ & inner edge of planetesimal disk & $0.01 \au$ & $0.01$-$0.1$ \\
		$a_\mathrm{pl,out}$ & outer edge of planetesimal disk & $3 \au$ & $3$-$25 \au$ \\
        $\langle e_0^2 \rangle^{1/2}$ & initial eccentricity dispersion of planetesimals & $10^{-3}$ & \\
        $\langle \sin^2 i_0 \rangle^{1/2}$ & initial inclination dispersion of planetesimals & $0.5\times10^{-3}$ & $10^{-4}$-$10^{-1}$ \\
		\hline
	\end{tabular}
	\caption{
	The various parameters used in this study (see text for details). 
    }
    \label{tb: parameters}
\end{table*}

We perform parameter studies where we change  $M_\mathrm{p}$, $R_\mathrm{p}$, $a_\mathrm{p}$, and  $e_\mathrm{p}$, which are also parameterised in \citet{Rodet+2023}.
We also perform parameter studies where we vary  $\sin i_\mathrm{p}$, and $\langle \sin^2 i_0 \rangle^{1/2}$. 
The values of the different parameters used in our simulations, as well as the range of parameters, are listed in Table ~\ref{tb: parameters}.

\subsection{Orbital evolution around an eccentric giant planet}\label{sec:standard}
  
\begin{figure}
  \begin{center}
    \includegraphics[width=80mm]{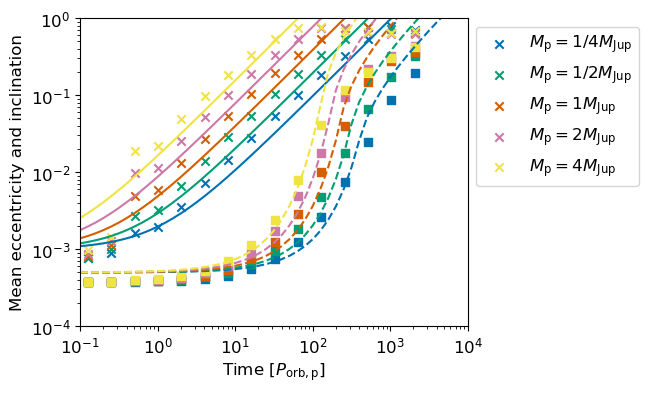}
    \caption{
    Time evolution of planetesimals' mean eccentricity (cross points) and inclination (square points).
    The different colors correspond to different assumed masses for the eccentric planet.
    We also plot eq.~(\ref{eq:TimeEvolution_Ecc}) with solid lines and eq.~(\ref{eq:TimeEvolution_Inc}) with dashed lines using $C_v=2$.
    Here, we show the results obtained in the standard cases with $a_\mathrm{p}=10 \au$.
    }
    \label{fig:time_orbit}
  \end{center}
\end{figure}
Figure~\ref{fig:time_orbit} shows the time evolution of the geometric mean eccentricity and inclination of planetesimals for various assumed planetary masses.
\textbf{
We present the case with a planet of $a_\mathrm{p}=1 \au$, $e_\mathrm{p}=0.99$, and $\sin i_\mathrm{p}=0$.
Eccentricity excitation occurs every planetary orbit and the mean eccentricity increases linearly with time.
The inclination excitation is much slower since the typical timescale of close encounters $\tau_\mathrm{close, 2D}$ is $20-100 P_\mathrm{orb,p}$ in these cases.
}
We also plot eq.~(\ref{eq:TimeEvolution_Ecc}) and eq.~(\ref{eq:TimeEvolution_Inc}) with $C_v=2$.
Equations (\ref{eq:TimeEvolution_Ecc}) and (\ref{eq:TimeEvolution_Inc}) reproduce the numerical results 
\textbf{
until the mean eccentricity gets closer to the planetary eccentricity $\left< e^2 \right>^{1/2} \sim e_\mathrm{p}$.
This is because secular resonances become effective when $\left< e^2 \right>^{1/2} \sim e_\mathrm{p}$. 
We will discuss the secular resonances' effect on planetesimals' orbital evolution in Sec.~\ref{sec: discussion_model}.
}

\subsection{Fate of planetesimals}\label{sec:fate_planetesimals}

\begin{figure}
  \begin{center}
    \includegraphics[width=80mm]{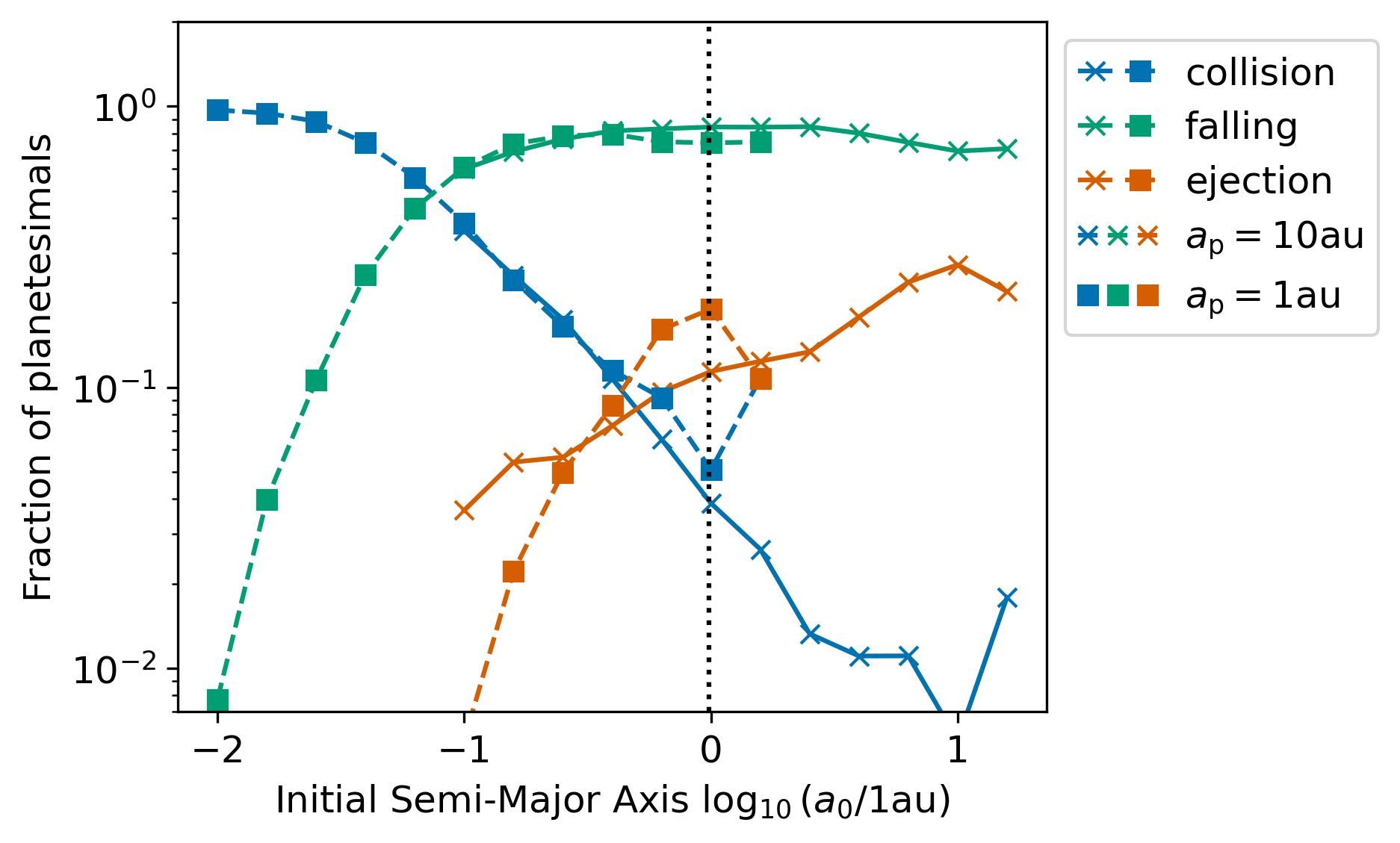}
    \caption{
    The fate of planetesimals around the eccentric planet of $e_\mathrm{p}=0.99$.
    The fractions of collision (blue), falling (green), and ejection (orange) are plotted as a function of the initial semi-major axis of planetesimals.
    We take the width of bins as 0.2 in logarithm space.
    Here, we show the results obtained in the standard cases with $a_\mathrm{p}=1 \au$ (dashed line) and $a_\mathrm{p}=10 \au$ (solid lines).
    The vertical dashed line shows the orbit of $\Theta=1$.
    }
    \label{fig:fate_standard}
  \end{center}
\end{figure}

Figure~\ref{fig:fate_standard} shows the final fate of planetesimals as a function of their initial semi-major axes.
The blue, green, and orange lines show the fraction of planetesimals that accrete on the eccentric planet $F_\mathrm{acc}$, fall to the central star $F_\mathrm{fall}$, and are ejected from the planetary system $F_\mathrm{eject}$, respectively. 
\textbf{
We present the case with a planet of $M_\mathrm{p}=1 M_\mathrm{Jup}$, $R_\mathrm{p}=1 R_\mathrm{Jup}$, $e_\mathrm{p}=0.99$, and $\sin i_\mathrm{p}=0$.
The solid and the dashed lines show the cases with $a_\mathrm{p}=10 \mathrm{au}$ and $1 \mathrm{au}$, respectively.
$F_\mathrm{acc}$ is larger in the inner orbit, which is consistent with the derived analytical model eq.~(\ref{eq:Pcol}).
}
For a Jupiter mass planet with Jupiter's radius, $f_\mathrm{a/c} = R_\mathrm{cap}/b_\mathrm{close}$ is $\sim0.76$, $0.082$ and $0.013$ at $r=0.01 \au$, $0.1 \au$ and $1\au$, respectively.
We find that $F_\mathrm{acc}$ is larger than $f_\mathrm{a/c}$, which means that planetesimals experience several close encounters with the eccentric planet before they are eliminated from the planetary system.

We also find that $F_\mathrm{fall}$ is increasing with the increasing semi-major axis of planetesimals.
Beyond a radial distance of $a\sim 0.1 \au$, planetesimals mostly fall into the central star.
On the other hand, $F_\mathrm{eject}$ increases in the outer disk region due to the weak gravitational bound from the central star.
However, the ejection from the planetary system is rare, and $F_\mathrm{eject}$ does not exceed $0.2$.
The vertical dotted line in Fig.~\ref{fig:fate_standard} shows the location of $\Theta=1$.
Our numerical results show that while a single close encounter at $\Theta<1$ is insufficient to eject planetesimals, a series of weak encounters accumulate perturbations on the planetesimal's orbit as shown in Fig.~\ref{fig:time_orbit}, which eventually leads to falling or ejection.

\begin{figure}
  \begin{center}
    \includegraphics[width=80mm]{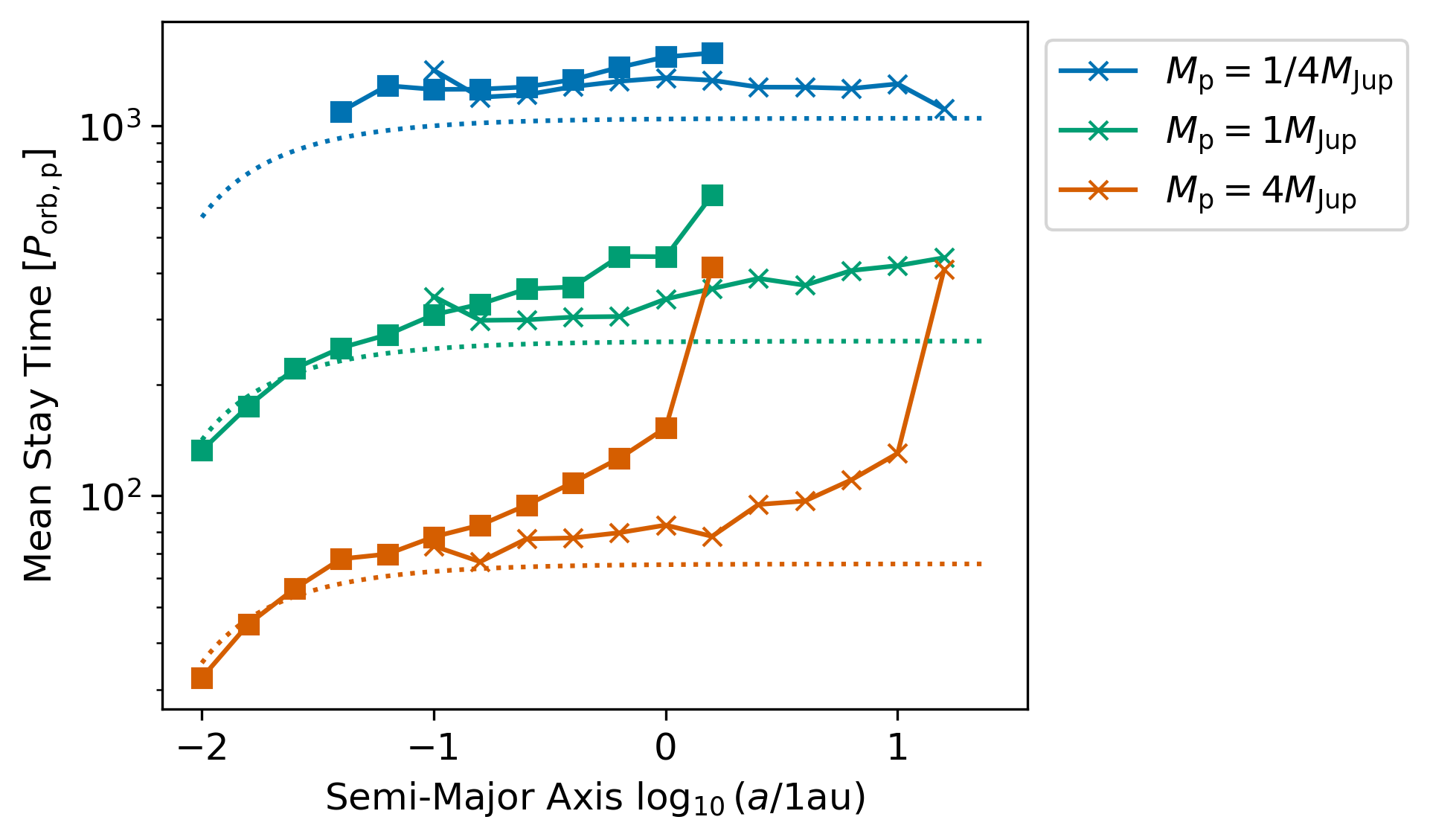}
    \caption{
    Averaged time before planetesimals fall to the central star or are ejected from the planetary system.
    Different colors correspond to the different masses of the eccentric planet.
    Square and cross plots show the cases with $a_\mathrm{p}=1 \au$ and $a_\mathrm{p}=10 \au$, respectively.
    The dotted lines show the analytical equation eq.~(\ref{eq:stay_time}) with a fitting parameter $C_\mathrm{v}=2$.
    }
    \label{fig:stay_time}
  \end{center}
\end{figure}

Figure~\ref{fig:stay_time} shows the geometric mean of the ejection time $\tau_\mathrm{eject}$ inferred in the numerical simulations.
\textbf{
We find that $\tau_\mathrm{eject}$ is shorter in the inner orbit because $e_\mathrm{fall}$ is smaller in this region. 
$\tau_\mathrm{eject}$ is shorter around the more massive planet since planetesimal excitation occurs more quickly.
}
The dotted lines show our analytical formula of $\tau_\mathrm{eject}$ given by eq.~(\ref{eq:stay_time}).
Around the aphelion of the eccentric planet, $\tau_\mathrm{stay}$ gets higher than eq.~(\ref{eq:stay_time}) because some planetesimals stay in mean motion resonances, which delay the eccentricity excitation of the planetesimals.
\textbf{
Sec.~\ref{sec: discussion_model} discusses this effect in more detail.
}

\subsection{Fitting the numerical results}

\begin{figure*}
  \begin{center}
    \includegraphics[width=160mm]{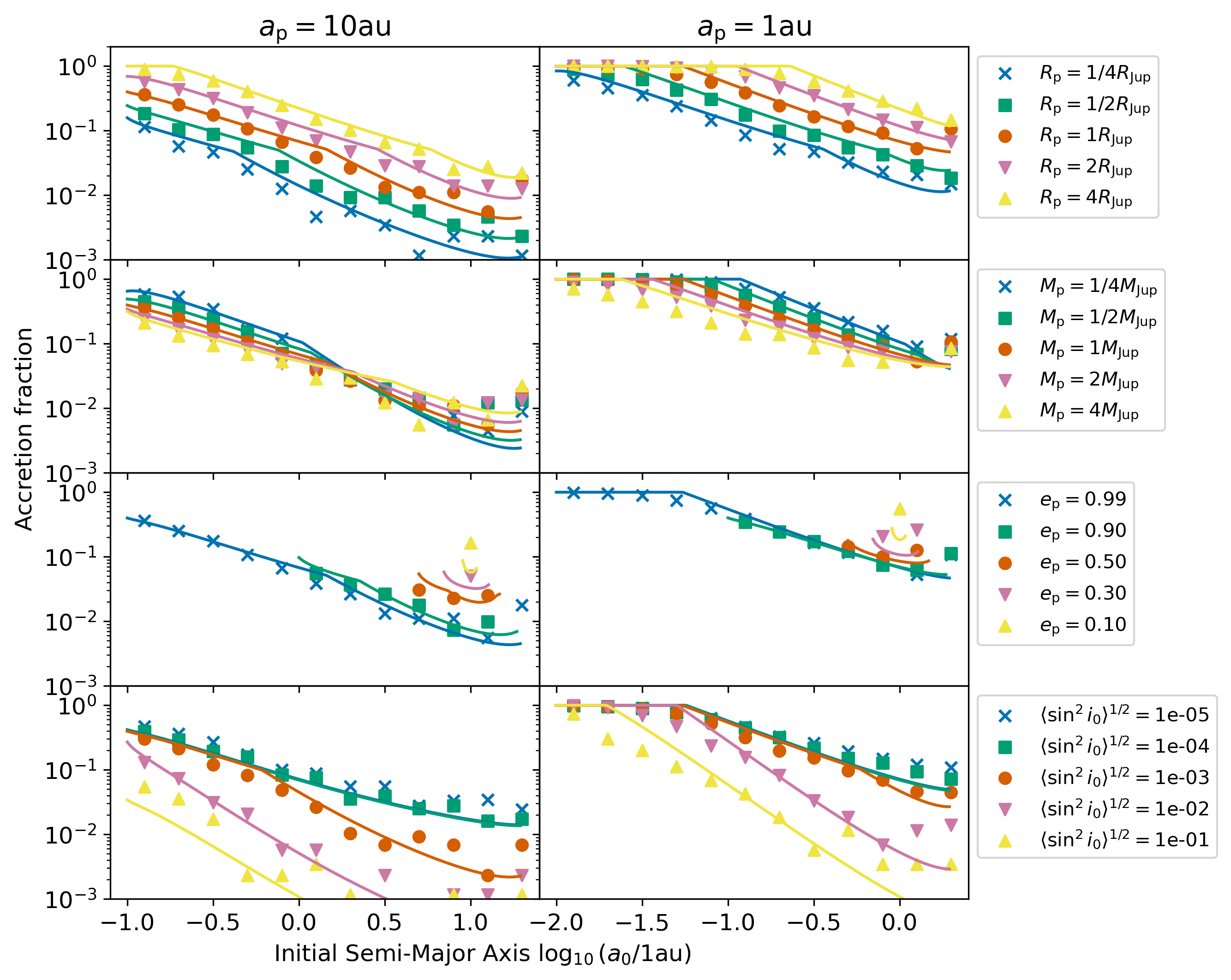}
    \caption{
    The fraction of collided planetesimals $F_\mathrm{acc}$ as a function of the initial semi-major axis of planetesimals.
    We plot the results of parameter studies for different assumed planetary radius $R_\mathrm{p}$, the planetary mass $M_\mathrm{p}$, the planetary eccentricity $e_\mathrm{p}$, and the initial disk thickness $\left \langle \sin^2 i_0 \right \rangle^{1/2}$ from top to bottom.
    The left and right columns show the cases with $a_\mathrm{p}=1\au$ and $a_\mathrm{p}=10\au$, respectively.
    The solid lines show the fitting formula given by eq.~(\ref{eq:Pcol}).
    }
    \label{fig: fitting}
  \end{center}
\end{figure*}

Figure~\ref{fig: fitting} shows the dependence of the accretion fraction of planetesimals $F_\mathrm{acc}$ (as a function of the initial semi-major axis of planetesimals) on the assumed planetary radius $R_\mathrm{p}$, mass $M_\mathrm{p}$, eccentricity $e_\mathrm{p}$, and the initial inclination of the planetesimal disk $\sin i_0$.
\textbf{
Parameters other than those labeled in the right boxes of each panel are set to the standard values we list in table~\ref{tb: parameters}.
Note that some parameter sets cover the unrealistic region, e.g., $R_\mathrm{p}=4 R_\mathrm{Jup}$ with $M_\mathrm{p} = 1 M_\mathrm{Jup}$. 
Still, we include these cases to assess our analytical model's validity.
}
The accretion probability increases with increasing $R_\mathrm{p}$ and decreases with increasing $M_\mathrm{p}$ and $e_\mathrm{p}$, which are consistent with the results by \citet{Rodet+2023}.
We also find that the accretion probability decreases when the initial semi-major axis of planetesimal $a_0$ and the initial inclination $\sin i_0$ increase because the geometrical picture of accretion changes from 2D to 3D. 

\textbf{
As shown in Fig.~\ref{fig:time_orbit}, the eccentricity excitation becomes slower than the prediction by eq.~(\ref{eq:TimeEvolution_Ecc}) as the eccentricity reaches values closer to the planetary eccentricity. 
This effect increases the probablity of close encounters before being eliminated from the planetary system. 
We introduce the following fitting parameters to eq.~(\ref{eq: Pcol_pp}) to account for this effect: 
\begin{eqnarray}
    P^{\prime\prime}_\mathrm{col} = \alpha P_\mathrm{2D} + \beta P_\mathrm{3D2D} + \gamma P_\mathrm{3D}. \label{eq: Pcol_pp_update}
\end{eqnarray}
We then fit the accretion probablity given by eq.~(\ref{eq:Pcol}) to the numerical results in logarithmic space.
}
The inferred fitting parameters are listed in table~\ref{tb:fitting_parameters} with the coefficient of determination being $0.951$. 
Equation~(\ref{eq:Pcol}) reproduces the numerical results except in the cases of low planetary eccentricity $e_\mathrm{p}\lesssim0.3$.
Our model underestimates the accretion probability of $e_\mathrm{p}\lesssim0.3$.
We discuss this point further in Sec.~\ref{sec: discussion_model}. 

\begin{table}
	\centering
	\begin{tabular}{ c|c } 
		\hline
		parameter & obtained value \\
		\hline
        $\alpha$   &  0.358 \\ 
        $\beta$   &  4.73 \\ 
        $\gamma$   &  7.81 \\ 
		\hline
	\end{tabular}
	\caption{
	Fitting parameters obtained from the numerical simulations
    }
    \label{tb:fitting_parameters}
\end{table}

\subsection{Dependence on the planetary inclination} 
\begin{figure*}
  \begin{center}
    \includegraphics[width=160mm]{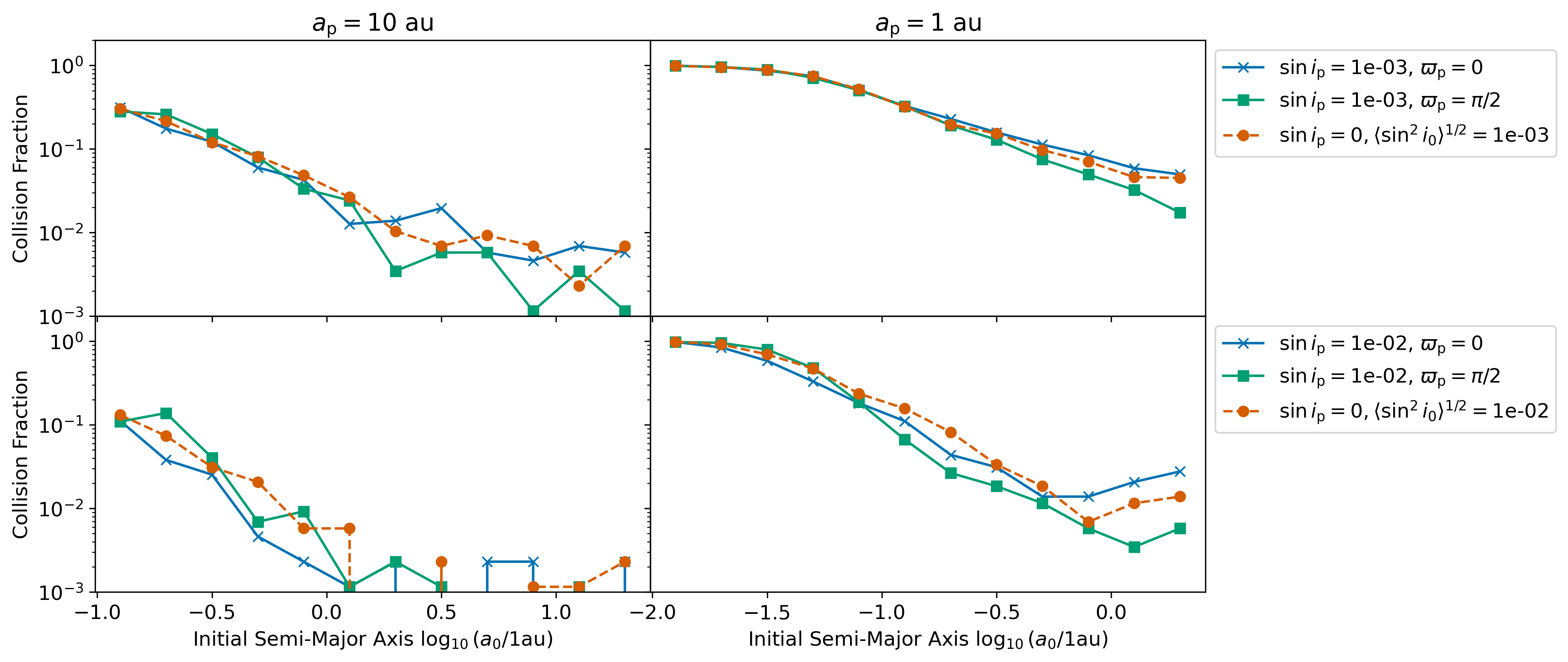}
    \caption{
    Same as Fig.~\ref{fig: fitting} but for different assumed planetary inclinations. 
    The blue and green solid lines show the cases with different longitudes of pericenter $\varpi_\mathrm{p}$.
    The orange dashed line shows the result for a non-inclined planet but with an inclination-excited planetesimal disk.}
    \label{fig:result_IncpIncpl}
  \end{center}
\end{figure*}

We considered the cases where the inclination of the eccentric planet is aligned to the planetesimal disk $\sin i_\mathrm{p}=0$.
If the eccentric planet is more inclined than the planetesimal disk $\sin i_\mathrm{p} > \left \langle \sin i \right \rangle$, the distance of the close encounter ${b_\mathrm{close}}^{\prime}$ is: 
\begin{eqnarray}
    C_\mathrm{v} \frac{\delta v}{v_\mathrm{K}} \frac{r \sin i_\mathrm{p}}{{b_\mathrm{close}}^{\prime}} &=& \sin i, \\
    {b_\mathrm{close}}^{\prime} &=& {b_\mathrm{close}} \sqrt{\frac{\sin i_\mathrm{p}}{\sin i}}.
\end{eqnarray}
The inclination evolution accelerates by $\sqrt{\sin i_\mathrm{p}/\sin i}$.
However, the planetesimal disk reaches $\sin i_\mathrm{p}<\left \langle \sin i \right \rangle$ within $\tau_\mathrm{eject}$, and the following orbital evolution would be the same as the cases with $\sin i_\mathrm{p}<\left \langle \sin i \right \rangle$.
Figure~\ref{fig:result_IncpIncpl} shows the numerical results with $\sin i_\mathrm{p}=10^{-3}$ and $10^{-2}$.
For comparison, we also plot the results with $\sin i_\mathrm{p}=0$, but with $\left \langle \sin^2 i_0 \right \rangle^{1/2}=10^{-3}$ and $10^{-2}$. 
We find that the collision fraction with different $\sin i_\mathrm{p}$ is consistent with the cases of corresponding $\left \langle \sin^2 i_0 \right \rangle^{1/2}$ within a factor of $\sim2$.  
We also change the longitude of the pericenter $\varpi_\mathrm{p}$ and find that the collision fraction is mostly independent of  $\varpi_\mathrm{p}$. 
For the cases with $\sin i_\mathrm{p} > \left \langle \sin i_0 \right \rangle$ the collision probability is given by $P_\mathrm{col}^\prime$ with $\sin i_\mathrm{p}$ substituted for $\left \langle \sin i_0 \right \rangle$.

\subsection{Effects of self-gravity}\label{sec:SelfGravity}

\begin{figure}
  \begin{center}
    \includegraphics[width=80mm]{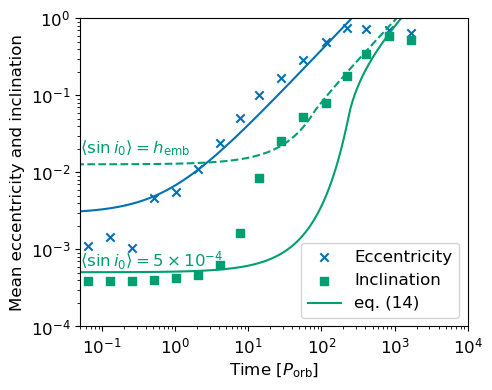}
    \caption{
    Same as Fig.~\ref{fig:time_orbit}, but using embryos instead of a planetesimal disk.
    The blue and green plots show the geometrical mean of eccentricities and inclinations of embryos.
    The blue solid line shows eq.~(\ref{eq:TimeEvolution_Ecc}) with $\langle e_0 \rangle=1\times10^{-3}$.
    The green solid and dashed lines show eq.~(\ref{eq:TimeEvolution_Inc}) with $\langle \sin i_0 \rangle = 5 \times 10^{-4}$ and $\langle \sin i_0 \rangle = h_\mathrm{emb}$, respectively. 
    Here, we show the results of $1M_\oplus$ embryos around $0.4 \au$.
    }
    \label{fig: time_orbit_Embryos}
  \end{center}
\end{figure}

\begin{figure}
  \begin{center}
    \includegraphics[width=80mm]{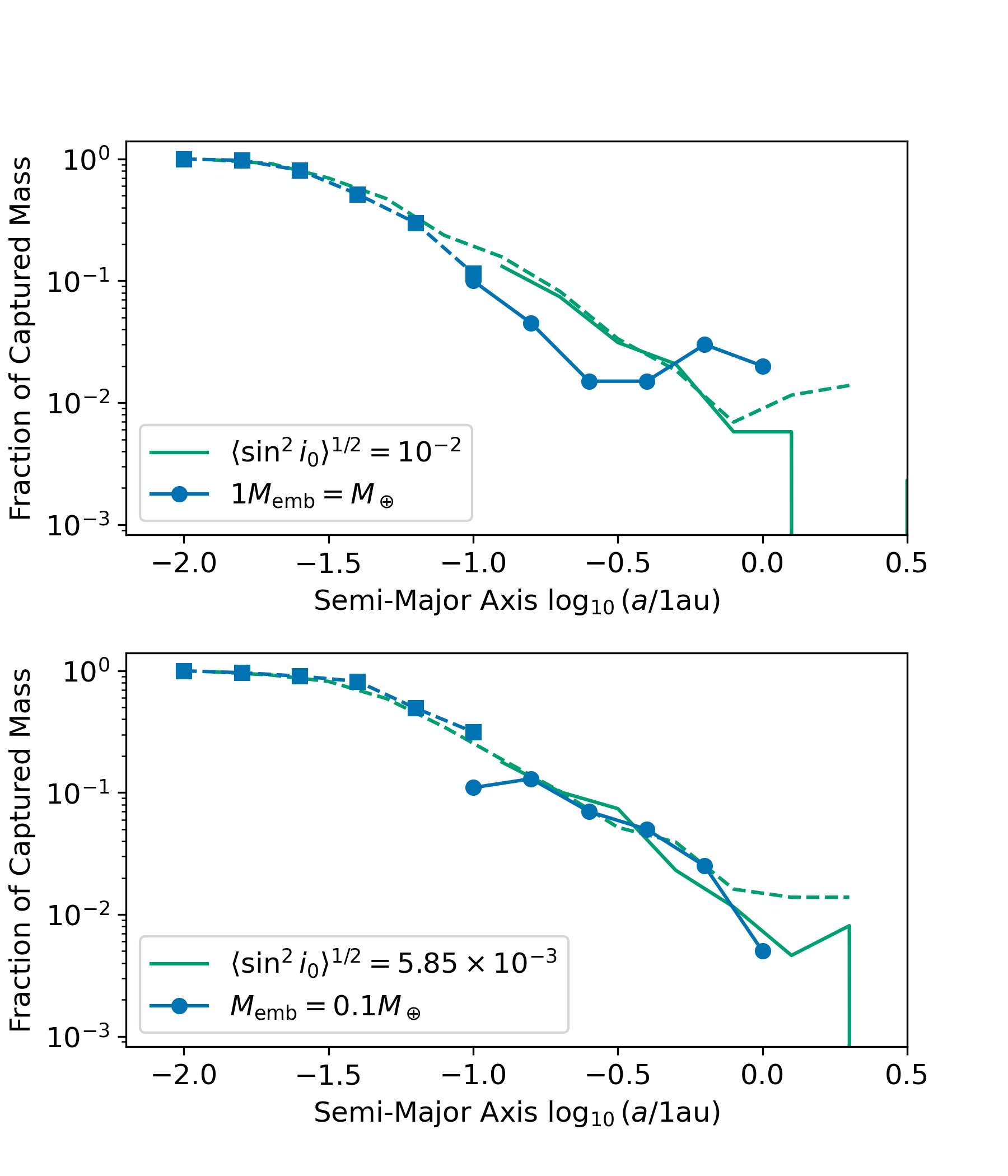}
    \caption{
    Same as Fig.~\ref{fig: fitting} but with the blue plots showing the results for embryos instead of the planetesimal disk. 
    We use five embryos for each simulation and sum up the collision fraction. 
    The upper panel shows the case with 1 $M_{\oplus}$ embryos, and the lower panel shows the case with 0.1 $M_{\oplus}$ embryos.
    For comparison, we also plot the results for the planetesimal disk with green lines. 
    The initial inclination of the planetesimal disk is equal to the reduced hill radius of embryos. 
    }
    \label{fig: embryos}
  \end{center}
\end{figure}

In the simulations discussed above, we treated planetesimals as test particles. 
We perform additional simulations using five equal-sized embryos instead of the planetesimal disk to investigate the effects of mutual gravitational interactions.
We set the mass of each embryo $M_\mathrm{emb}$ to $1 M_\oplus$ and $0.1 M_\oplus$.
The embryos' mean density is set to $5.5 \mathrm{g/cm^3}$, and the initial semi-major axis of the innermost embryo is an input parameter $a_\mathrm{emb,0}$.
The remaining embryos are distributed with the radial orbital separation of 10 mutual Hill radii.
The eccentricity and inclination are given by the Rayleigh distribution with $\left \langle e_0^2 \right \rangle^{1/2} = 2 \left \langle \sin^2 i_0 \right \rangle^{1/2} = 10^{-3}$, and the orbital angles are set randomly.
We perform simulations with 40 different seeds for each $a_\mathrm{emb,0}$ and assume perfect mergers for collisions.

Due to the mutual gravitational interactions, the eccentricity and inclination are excited to $\sim h_\mathrm{emb}$, which is a reduced mutual Hill radius $h_\mathrm{emb}= (2 M_\mathrm{emb}/3 M_\mathrm{s})^{1/3}$.
After the embryos' inclination reaches $\left \langle \sin i \right \rangle \sim h_\mathrm{emb}$, it evolves similarly to the planetesimals.
Figure~\ref{fig: time_orbit_Embryos} shows the time evolution of the geometrical mean of eccentricities and inclinations of embryos.
The embryos' inclination increases with time more rapidly than planetesimals in the early evolution stage $t \lesssim 10 P_\mathrm{orb,p}$ because the mutual interactions between the embryos dominate the inclination evolution.
After that stage, the inclination evolution follows that expected from eq.~(\ref{eq:TimeEvolution_Inc}) with $\left \langle \sin i_0 \right \rangle \sim h_\mathrm{emb}$.

Figure~\ref{fig: embryos} shows the fraction of collisions to the planet $F_\mathrm{col,emb}$ as a function of the semi-major axis. $F_\mathrm{col,emb}=M_\mathrm{col, tot}/M_\mathrm{emb, tot}$ where $M_\mathrm{col, tot}$ is the total mass of the embryos colliding with the eccentric planet. $M_\mathrm{emb,tot}=200 M_\mathrm{emb}$ is the total mass of embryos used for each $a_\mathrm{emb,0}$.
We also plot the accretion fraction $F_\mathrm{acc}$ obtained in the simulations using the planetesimal disk with $\left \langle \sin^2 i_0 \right \rangle^{1/2} = h_\mathrm{emb}$.
We find that the collision fraction $F_\mathrm{col,emb}$ is comparable to the accretion fraction $F_\mathrm{acc}$.
Therefore, we can conclude that the collision probability of embryos is also given by $P_\mathrm{col}^\prime$ with $\left \langle \sin i_0 \right \rangle=h_\mathrm{emb}$.

\section{The effect of the planetary eccentricity excitation}
\label{sec: EccentricityEvolution}
In the previous section, we inferred the accretion probability of a giant planet with a fixed orbit without considering the planet's orbital evolution. 
Here, we investigate how the planet's eccentricity evolution affects the inferred accretion probability. 

\subsection{Eccentricity excitation}\label{sec:method_EccExcite}
\textbf{
We consider a case where multiple planets orbit the exterior of the planetesimal disk.
The planets initiate orbital instability and increase eccentricities, and then the pericenter of the innermost planet starts to cross the planetesimals' orbit.
We assume that the gravitational interaction with the innermost planet is the dominant perturber onto the planetesimal disk and neglects the gravity from the outer planets since the gravitational force from the outer planets is much smaller than that from the innermost planet. 
}

We start the N-body simulations with $e_\mathrm{p}=0$ and increase the planetary eccentricity by $1/\tau_\mathrm{ecc}$ every orbit where $\tau_\mathrm{ecc}$ is the excitation timescale.
There are various mechanisms for eccentricity excitation, and the excitation timescale differs in each mechanism. 
Typical timescales of the Kozai-Lidov resonance and secular interactions could be $10^5 P_\mathrm{orb,p}$ or longer \citep[e.g.][]{Dawson+2018}.
On the other hand, planet-planet scattering can increase the eccentricity more quickly.
The eccentricity excitation kicked by a single encounter is given by: 
\begin{eqnarray}
    \Delta e &\lesssim& \frac{v_\mathrm{esc}}{v_\mathrm{K}}, \\
            &=& 0.32 \left( \frac{M_\mathrm{p}}{0.5M_\mathrm{Jup}} \right)^{1/2} \left( \frac{R_\mathrm{p}}{2 R_\mathrm{Jup}} \right)^{-1/2} \left( \frac{a_\mathrm{p}}{0.1 \mathrm{au}} \right)^{1/2} \left( \frac{M_\mathrm{s}}{M_\oplus} \right)^{-1/2}.
\end{eqnarray}
N-body simulations have shown that planet-planet scattering can excite the planetary eccentricity as quickly as $10^3 P_\mathrm{orb,p}$ at most \citep{Carrera+2019, Pu+2021, Garzon+2022}.
We consider the following values of  $\tau_\mathrm{ecc}=10^3, 10^4, 10^5 P_\mathrm{orb,p}$ to account for the various excitation mechanisms. 

While the secular interactions keep the semi-major axis constant \citep{Teyssandier+2019}, the planet-planet scattering changes other orbital parameters, such as the planetary semi-major axis and inclination \citep{Carrera+2019, Pu+2021, Garzon+2022}. 
However, here, we increase the eccentricity by keeping the semi-major axis and the inclination constant to investigate how the time evolution of the planetary pericenter affects the accretion probability.

\subsection{Case with $\tau_\mathrm{ecc}=10^3 P_\mathrm{orb,p}$}\label{sec: OrbitalEvolutionBOC_3}

\begin{figure*}
  \begin{center}
    \includegraphics[width=150mm]{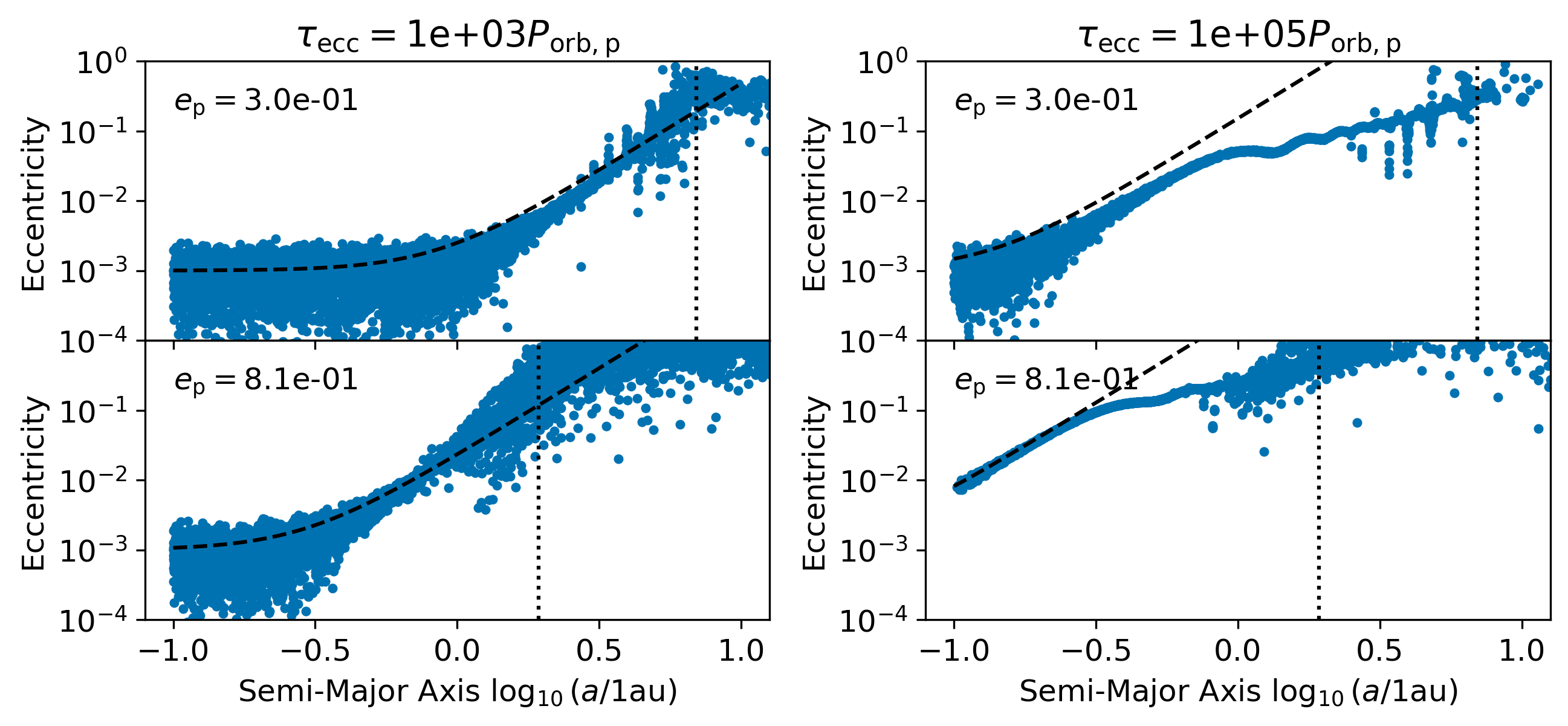}
    \caption{
    Snapshots of the planetesimal disk when the planet's eccentricity increases with time.
    The left and right columns show the cases with $\tau_\mathrm{ecc}=10^3 P_\mathrm{orb,p}$ and $10^3 P_\mathrm{orb,p}$, respectively.
    The upper and lower panels show the snapshots when the planetary eccentricity is $e_\mathrm{p}=0.3$ and $0.81$, respectively.
    The vertical dotted lines are the pericenter and apocenter of the planet.
    The dashed lines are the analytical models of the planetesimal eccentricity given by eq.~(\ref{eq:ecc_PassingStarEvolv}).
    When $\tau_\mathrm{ecc}=10^5 P_\mathrm{orb,p}$, the planetesimal eccentricity is smaller than that given by eq.~(\ref{eq:ecc_PassingStarEvolv}) because the resonances suppress the linear increase of the eccentricity.
    }
    \label{fig: Snapshots_psdEccp}
  \end{center}
\end{figure*}

\textbf{
First, we show the result with a planet whose eccentricity increases as rapidly as $\tau_\mathrm{ecc}=10^3 P_\mathrm{orb,p}$.
The left panels in Fig.~\ref{fig: Snapshots_psdEccp} show snapshots of the planetesimal disk. 
The upper and lower panels show when the planet's eccentricity becomes $0.30$ and $0.81$, respectively.
The vertical dotted lines are the pericenter of the eccentric planet.
We find that planetesimals are excited to the eccentric orbit before the planet's pericenter reaches the planetesimals' orbit.
This is because of the indirect forces of the eccentric planet.
}
\cite{Kobayashi+2001} showed that the eccentric object can perturb the inner planetesimal disk without orbital crossing.
The eccentricity perturbation put by a passing object is given as \citep{Kobayashi+2001}:
\begin{eqnarray}
    \Delta e_\mathrm{pass} &\simeq& A \left( \frac{a}{q_\mathrm{p}} \right)^{5/2},  \label{eq:ecc_PassingStar} \\
    A &=& \frac{15 \pi}{32 \sqrt{2}}  \frac{M_\mathrm{p}/M_\mathrm{s}}{\sqrt{1+M_\mathrm{p}/M_\mathrm{s}}} 
    \left[ \cos^2 \omega_\mathrm{p} \left(1-\frac{5}{4} \sin^2 i_\mathrm{p}\right)^2 \right. \nonumber \\
    && \left. + \sin^2\omega_\mathrm{p} \cos^2 i_\mathrm{p} \left(1-\frac{15}{4} \sin^2 i_\mathrm{p}\right)^2 \right]^{1/2}, \label{eq:ecc_PassingStar2}
\end{eqnarray} 
where $q_\mathrm{p}$ is the pericenter of the planet.
We adopt eq.~(\ref{eq:ecc_PassingStar}) to the eccentric planet and integrate it along the planetary eccentricity evolution.  
We obtain the eccentricity of the planetesimal disk, which is given by: 
\begin{eqnarray}
    \left \langle e \right \rangle &\simeq& \int_0^{t} \Delta e_\mathrm{pass} \mathrm{d} t^\prime = \int_0^{e_\mathrm{p}} \Delta e_\mathrm{pass} \frac{\tau_\mathrm{ecc}}{P_\mathrm{orb,p}} \mathrm{d} e_\mathrm{p}^\prime, \nonumber \\
    &=& \left \langle e_0 \right \rangle +\frac{2}{3} A \tau_\mathrm{ecc} \left(\frac{a}{a_\mathrm{p}}\right)^{5/2} \left[ \left( 1- e_\mathrm{p} \right)^{-3/2} -1 \right]. \label{eq:ecc_PassingStarEvolv}
\end{eqnarray}
\textbf{
We plot eq.~(\ref{eq:ecc_PassingStarEvolv}) with the dashed lines in Fig.~\ref{fig: Snapshots_psdEccp}.
}
We find that eq.~(\ref{eq:ecc_PassingStarEvolv}) reproduces well the eccentricity evolution of the planetesimals.

\begin{figure}
  \begin{center}
    \includegraphics[width=80mm]{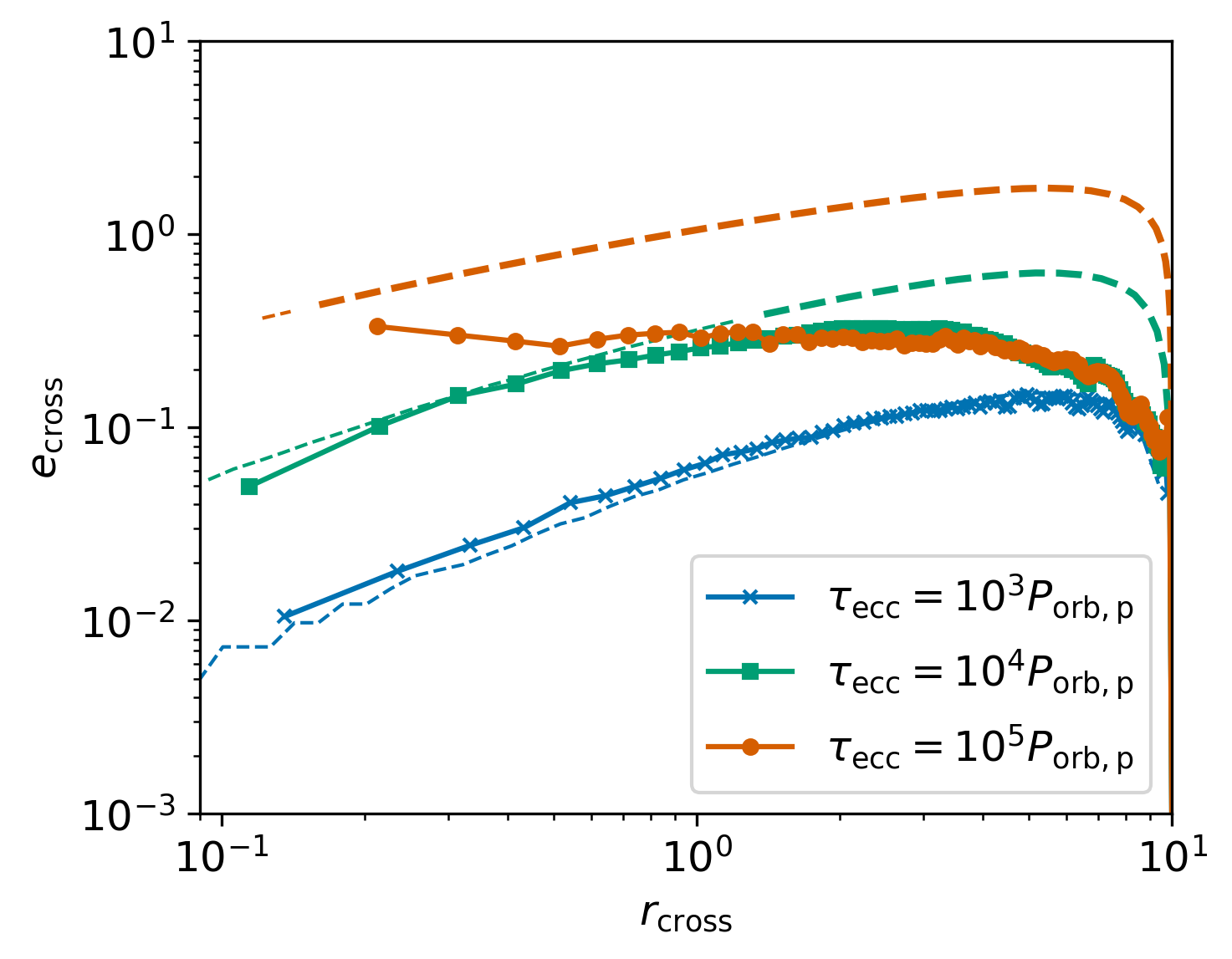}
    \caption{
    $r_\mathrm{cross}=a_\mathrm{p}(1-e_\mathrm{cross,p})$ vs. $e_\mathrm{cross}$ obtained by the N-body simulations.
    $e_\mathrm{cross}$ and $e_\mathrm{cross,p}$ are calculated by taking the average of eccentricities of planetesimals whose orbits are just before crossing to the planet's orbit.
    The dashed lines show $e_\mathrm{cross}$ obtained from eq.~(\ref{eq:ecc_PassingStarEvolv}) and eq.~(\ref{eq:orbits_crossed}).
    The thick dashed lines mean eq.~(\ref{eq: a_crit}) is achieved. 
    When eq.~(\ref{eq: a_crit}) is achieved, $e_\mathrm{cross}$ becomes smaller than the analytical models predict.
    As indicated in the legend, the different colors correspond to the different eccentricity excitation timescales $\tau_\mathrm{ecc}$.}
    \label{fig: Ecross}
  \end{center}
\end{figure}

The condition where the planet crosses the orbit of planetesimals is given by: 
\begin{eqnarray}    
    a \left(1+e\right) = a_\mathrm{p} \left( 1-e_\mathrm{p} \right). \label{eq:orbits_crossed}
\end{eqnarray}
By solving eq.~(\ref{eq:ecc_PassingStarEvolv}) and eq.~(\ref{eq:orbits_crossed}) numerically, we can obtain the eccentricity of planetesimals and the planet when the orbits are crossed $e_\mathrm{cross}$, and $e_\mathrm{cross,p}$, separately.
\textbf{
Figure~\ref{fig: Ecross} shows $e_\mathrm{cross}$ as a function of $r_\mathrm{cross}=a (1+e_\mathrm{cross})$.
The plots with solid lines are numerical results inferred from the N-body simulations. 
}
We obtain $e_\mathrm{cross}$ by calculating the mean eccentricity of planetesimals whose orbits are just before crossing to the planet's orbit: $0.8<a(1+e)/a_\mathrm{p}(1-e_\mathrm{p})<0.9$.
The dashed lines show the analytical prediction of $e_\mathrm{cross}$ obtained from eq.~(\ref{eq:ecc_PassingStarEvolv}) and eq.~(\ref{eq:orbits_crossed}). 
\textbf{
When $\tau_\mathrm{ecc}=10^3 P_\mathrm{orb,p}$ (blue lines), the analytical model reproduces the numerical result well, and $e_\mathrm{cross}$ is smaller than $0.1$.
Therefore, planetesimals have many chances to have close encounters after the orbits are crossed.
}

\begin{figure}
  \begin{center}
    \includegraphics[width=80mm]{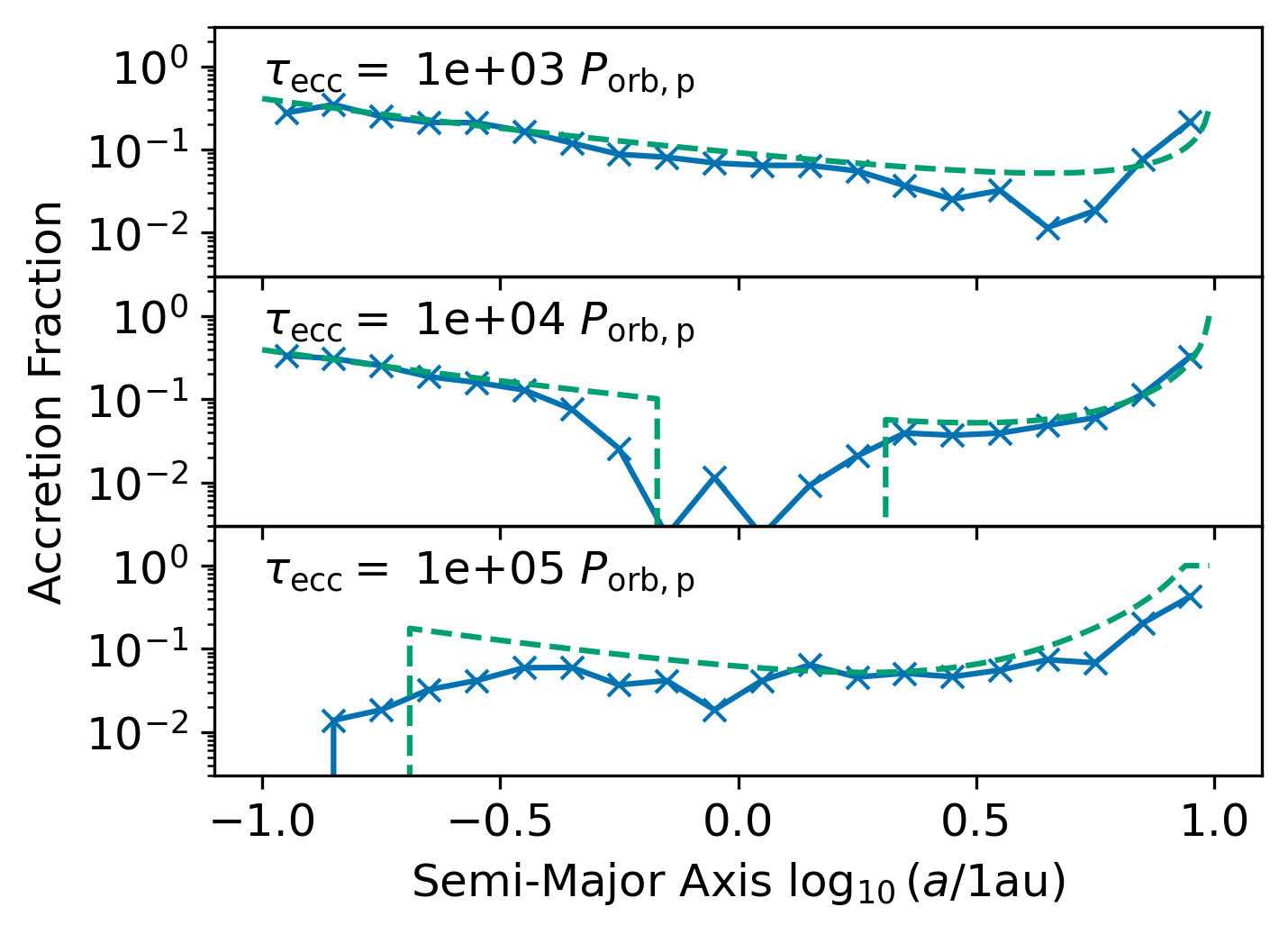}
    \caption{
    Same as Fig.~\ref{fig: fitting} but with a planet whose eccentricity increases with a given timescale $\tau_\mathrm{ecc}$.
    The blue solid lines show the numerical results and the green dashed lines show eq.~(\ref{eq: accretion_probability}).
    }
    \label{fig:collision_psdEccp}
  \end{center}
\end{figure}

Figure~\ref{fig:collision_psdEccp} shows the accretion fraction of planetesimals $F_\mathrm{acc}$ vs.~the initial semi-major axis of planetesimals.
$F_\mathrm{acc}$ exceeds $10\%$ in the inner disk $a_0<1 \mathrm{au}$ and around the initial orbit of the planet $a_0\sim10\mathrm{au}$.
When the planet's eccentricity is small, the gravitational focusing works effectively, and $F_\mathrm{acc}$ can be high. 
As the planet is excited to the eccentric orbit, its pericenter reaches the inner orbit. 
The planet accretes planetesimals there, but the accretion probability decreases because of the weaker gravitational focusing effect.
This effect leads to a peak on $F_\mathrm{acc}$ around $a_0\sim a_\mathrm{p}$.
On the other hand, $F_\mathrm{acc}$ gets higher in the further inner disk, too, because $R_\mathrm{p}/b_\mathrm{close}$ is larger there.
\textbf{
The accretion probability can be estimated with eq.~(\ref{eq:Pcol}) by substituting $a_0=r_\mathrm{cross}$ and $e_\mathrm{p}=e_\mathrm{p,cross}$.
The green dashed line in Fig.~\ref{fig:collision_psdEccp} shows eq.~(\ref{eq:Pcol}).
Equation~(\ref{eq:Pcol}) reproduces the numerical results well when $\tau_\mathrm{ecc}=10^3 P_\mathrm{orb,p}$.
}

\subsection{Cases with larger $\tau_\mathrm{ecc}$}\label{sec: OrbitalEvolutionBOC}
\textbf{
As the eccentricity excitation of the planet gets slower, $e_\mathrm{cross}$ becomes larger since the perturbations from the eccentric planet accumulate more before the orbits are crossed.
The analytical model of $e_\mathrm{cross}$ given by eq.~(\ref{eq:ecc_PassingStarEvolv}) and eq.~(\ref{eq:orbits_crossed}) predicts that $e_\mathrm{cross}$ becomes as large as $1$ when $\tau_\mathrm{ecc}=10^5 P_\mathrm{orb,p}$ (the orange dashed line in fig.~\ref{fig: Ecross}).
This means that planetesimals could be ejected from the planetary system before the orbits of the planetesimals and the planet are crossed, which should decrease the accretion rate.
However, numerical results show that $e_\mathrm{cross}$ is much smaller than the analytical model predicts (the orange plots in Fig.~\ref{fig: Ecross}), and the planetesimals can accrete to the eccentric planet as shown in the lower panel of Fig.~\ref{fig:collision_psdEccp}.
}
This is because of the effect of secular resonances.
\citet{Kobayashi+2001} found that resonance effects appear around $r/q_\mathrm{p}\gtrsim0.2$.
This resonance effects suppress the eccentricity excitation
\textbf{
and $e_\mathrm{cross}$ becomes smaller than the analytical model predicts.
Due to the effective secular resonances, the planetesimal disk can accrete onto the eccentric planet even if the eccentricity excitation of the planet is as slow as $\tau_\mathrm{ecc}=10^5 P_\mathrm{orb,p}$.
}

\textbf{
On the other hand, the accretion fraction is reduced around $1 \mathrm{au}$ and opens a gap in $F_\mathrm{acc}$ when $\tau_\mathrm{ecc}=10^4 P_\mathrm{orb,p}$.
In the region inner than $1 \mathrm{au}$, $e_\mathrm{cross}$ is small, and the planetesimals can interact with the eccentric planet after the orbits are crossed.
However, $e_\mathrm{cross}$ is increasing with $a_\mathrm{0}$, and once $e_\mathrm{cross}$ exceeds $\sim0.3$, $F_\mathrm{acc}$ drops to 0.
We define a critical value $e_\mathrm{cross, crit}$ as the upper limit for planetesimals to accrete onto the eccentric planet.
Planetesimal accretion occurs in the region where
\begin{align}
    e_\mathrm{cross} < e_\mathrm{cross, crit}. \label{eq: e_cross_crit}
\end{align}
}

\textbf{
If the secular resonances become effective, the planetesimals can accrete onto the eccentric planet even if the analytical model given by eq.~(\ref{eq:ecc_PassingStarEvolv}) and eq.~(\ref{eq:orbits_crossed}) does not achieve eq.~(\ref{eq: e_cross_crit}).
For the resonances to be effective, the pericenter of the planet needs to move slower than $\sim 1/\tau_\mathrm{res}$, where $\tau_\mathrm{res}$ is the typical timescale of the secular resonances.
}
By analyzing the eccentricity evolution of each planetesimal, we find that the typical timescale of the resonance $\tau_\mathrm{res}$ is $\sim10^3~P_\mathrm{orb,p}$ (see details in Appendix~\ref{app: resonances}). 
\textbf{
The region where the resonances become effective is
\begin{eqnarray}
    \left| \frac{a_0}{\dot{q}_\mathrm{p}} \right| &> \tau_\mathrm{res}, \\
    \frac{\tau_\mathrm{res}}{\tau_\mathrm{ecc}} a_\mathrm{p} &< a_0. \label{eq: a_crit}
\end{eqnarray}
Neither of eq.~(\ref{eq: e_cross_crit}) and eq.~(\ref{eq: a_crit}) are not achieved around $1 \mathrm{au}$ when $\tau_\mathrm{ecc}=10^4 P_\mathrm{orb,p}$.
Therefore, $F_\mathrm{acc}$ has a gap around $1 \mathrm{au}$.
Because of the same reason, $F_\mathrm{acc}$ around $0.1 \mathrm{au}$ decreases when $\tau_\mathrm{ecc}=10^5 P_\mathrm{orb,p}$.
}

\textbf{
Including the above effects, we finally 
}
model the accretion probability of planetesimals by
\begin{eqnarray}\label{eq: accretion_probability}
    P_\mathrm{acc} =
    \begin{cases}
          P_\mathrm{acc}^\prime
        &~\text{~for~eq.~(\ref{eq: e_cross_crit})~or~eq.~(\ref{eq: a_crit})}, \\
         0 
        &\text{~other~},
    \end{cases}
\end{eqnarray}
We set $\tau_\mathrm{res}=2\times10^3 P_\mathrm{orb}$ and $e_\mathrm{cross,crit}=0.3$ from the numerical results. 
$r_\mathrm{cross}$, $e_\mathrm{cross}$ and $e_\mathrm{p,cross}$ are obtained with eq. (44) and eq. (45).
Other parameters such as $M_\mathrm{p}$, $R_\mathrm{p}$, $a_\mathrm{p}$, $\sin i_\mathrm{p}$ and $\left \langle \sin^2 i_\mathrm{0} \right \rangle^{1/2}$ are input parameters.
\textbf{
The green dashed lines in Fig.~\ref{fig:collision_psdEccp} show the accretion probability given by eq.~(\ref{eq: accretion_probability}).
}

\section{Discussion}
\label{sec: Discussion}
\subsection{The limitations of our orbital evolution model}\label{sec: discussion_model} 

\begin{figure}
  \begin{center}
    \includegraphics[width=80mm]{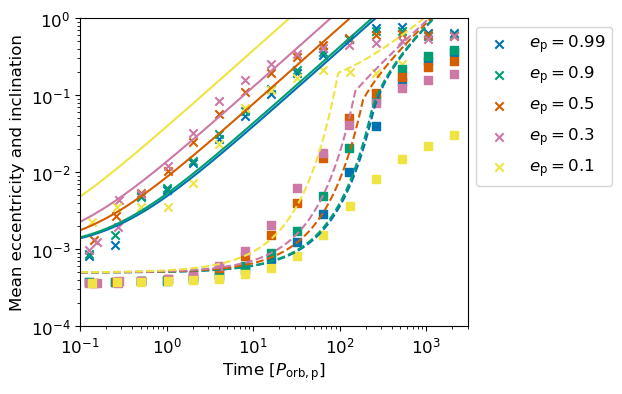}
    \caption{
    Same as Fig.~\ref{fig:time_orbit}, but for the results of parameter study of the planetary eccentricity $e_\mathrm{p}$.
    }
    \label{fig: time_orbits_psEccp}
  \end{center}
\end{figure}

In sec.~\ref{sec: CollisionToPlanet_ANL}, we model the orbital evolution of planetesimals around an eccentric planet.
A few effects have been neglected for simplicity, although some of them could be important under certain conditions. 
Figure~\ref{fig: time_orbits_psEccp} shows the time evolution of the mean eccentricities and the mean inclinations of planetesimals for different planetary eccentricity cases. 
The figure clearly shows that the excitation rates of $\left \langle e \right \rangle$ (cross dots) and $\left \langle \sin i \right \rangle$ (square dots) become slower than predicted by our model (solid and dashed lines) once $\left \langle e \right \rangle$ exceeds the planetary eccentricity.
Also, our model is invalid when $e_\mathrm{p}=0.1$. 
Here, we discuss our model assumptions and the limitations of our orbital evolution model. 

\subsubsection{A-particle-in-a-box approximation}
In a-particle-in-a-box model, we assume that the perturbation is negligibly small so that the trajectory of the planetesimal is unchanged. 
In the case of a close encounter where the trajectory of the perturbed particle is deflected during the passage of the eccentric planet, a-particle-in-a-box approximation is not valid.  
By substituting $\sim v_\mathrm{rel}$ to $\delta v$, we estimate the minimum relative distance available in a-particle-in-a-box approximation $b_\mathrm{min}$ as: 
\begin{eqnarray}
    b_\mathrm{min} \sim \frac{2 C_\mathrm{v}}{{e_\mathrm{p}}^2} \frac{M_\mathrm{p}}{M_\mathrm{s}} a.
\end{eqnarray}
Our model is appropriate when: 
\begin{eqnarray}
    b_\mathrm{close} &>& b_\mathrm{min}, \\
    e_\mathrm{p} &>& 0.156 \left( \frac{M_\mathrm{p}}{M_\mathrm{Jup}} \right)^{1/3} \left( \frac{M_\mathrm{s}}{M_\odot} \right)^{1/3} \label{eq: bmin_ecc}
\end{eqnarray}
with $C_\mathrm{v}=2$.
If the planetary eccentricity is lower than eq.~(\ref{eq: bmin_ecc}), the velocity kick of close encounter becomes $\sim \sqrt{\mathcal{G} M_\mathrm{p}/b}$ and $b_\mathrm{close}$ is redefined as: 
\begin{eqnarray}
    {b_\mathrm{close}}^\prime = \min \left( b_\mathrm{close}, {C_\mathrm{v}}^{2/3} \left( \frac{2 M_\mathrm{p}}{M_\mathrm{s}} \right)^{1/3} \right).
\end{eqnarray}
Since ${b_\mathrm{close}}^\prime$ is smaller than $b_\mathrm{close}$, the close encounter timescale gets longer compared to the one we assumed. 
Equation~(\ref{eq:TimeEvolution_Inc}) overestimates the time evolution of the inclination of planetesimals.
However, this effect is minor because the difference between ${b_\mathrm{close}}$ and ${b_\mathrm{close}}^\prime$ is only a few \% when $e_\mathrm{p}=0.1$. 


\subsubsection{High eccentricity effects on the eccentricity and inclination kick}

\begin{figure*}
  \begin{center}
    \includegraphics[width=160mm]{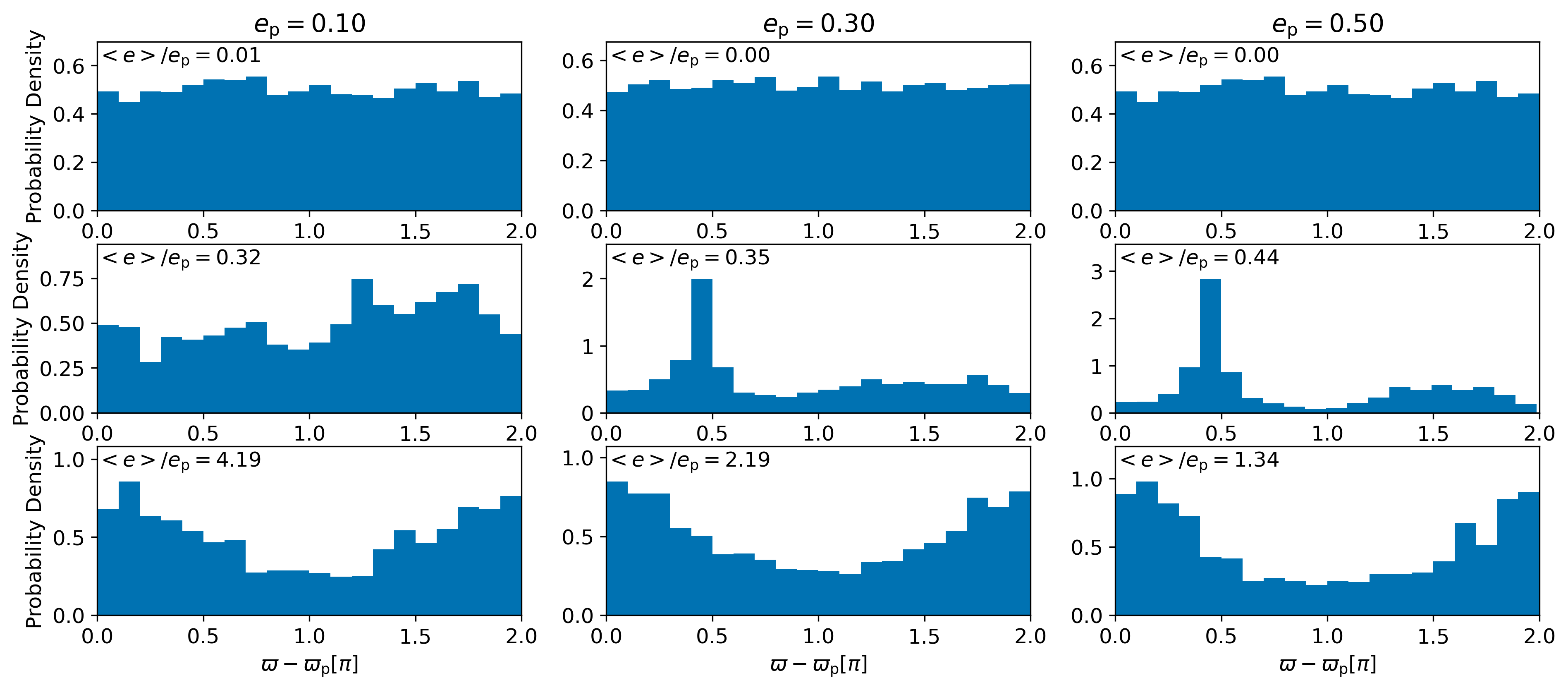}
    \caption{
    Distribution of $\varpi-\varpi_\mathrm{p}$ around an eccentric planet.
    We plot the cases of $e_\mathrm{p}=0.1$, $0.3$, and $0.5$ in the left, center, and right columns, respectively.
    The upper, middle, and lower panels show different times.
    The mean eccentricities of planetesimals divided by the planetary eccentricity are indicated in the upper left corner of each panel.
    Once $\left\langle e \right\rangle$ exceeds $e_\mathrm{p}$, the planetesimals' orbits start to align to the planetary orbit.
    }
    \label{fig: EccJupiter_Resonances}
  \end{center}
\end{figure*}

To transform a given velocity kick to perturbations on eccentricity and inclination, we use 
the approximated relations of 
$\delta e \sim \delta v/\delta v_\mathrm{K}$ and $\delta \sin i \sim \delta v_\perp/\delta v_\mathrm{K}$, where we neglect the eccentricity terms. 
The exact formulae are given by: 
\begin{eqnarray}
    \delta e &=& \sqrt{1-e^2} \left\{ \sin f \frac{\delta v_\mathrm{R}}{v_\mathrm{K}} +\left( \cos f + \cos E \right) \frac{\delta v_\mathrm{T}}{v_\mathrm{K}} \right\}, \\
    \delta \sin i &=& \sqrt{1-e^2} \frac{\cos \left( \omega + f \right)}{1+e \cos f} \frac{\delta v_\mathrm{Z}}{v_\mathrm{K}},
\end{eqnarray}
where $E$ is the eccentric anomaly and $\omega$ is the argument of perihelion.
The term of $\sqrt{1-e^2}$ makes the eccentricity and inclination evolution slower as $e$ increases.

We also assume that orbits between the planetesimals and the planet are crossed, and velocity kicks are put on the planetesimals' orbit independent of $f$ in every planetary orbit.
However, this geometrical picture does not apply when $\left\langle e \right\rangle > e_\mathrm{p}$.
Figure~\ref{fig: EccJupiter_Resonances} shows the histogram of the differences between the pericenters of the planetesimals $\varpi$ and the planet $\varpi_\mathrm{p}$.
Each column shows the cases of $e_\mathrm{p}=0.1$, $0.3$, and $0.5$.
We plot the mean eccentricity of planetesimals divided by the planetary eccentricity in each panel.
We find that $\varpi - \varpi_\mathrm{p}$ are distributed uniformly or around $\sim \pi/2$ if $\left\langle e \right\rangle < e_\mathrm{p}$.
In this case, most of the planetesimals cross the planet's orbit.
However, secular resonances align the pericenters between the planetesimals and the planet when $\left\langle e \right\rangle > e_\mathrm{p}$.
The alignment of the pericenters detaches the planetesimal's orbits from the planetary orbit.
Since the velocity kicks are weaker after the detachment, the orbital excitation gets slower once the planetesimals' eccentricities exceed the planetary eccentricity. 

\subsubsection{Co-rotation resonance}

When $e_\mathrm{p}=0.1$, the mean eccentricities and inclinations of planetesimals are smaller than the other cases even when $\left\langle e \right\rangle < e_\mathrm{p}$.
This is because many planetesimals are trapped in the co-rotation resonance 
\textbf{
(see snapshots in appendix ~\ref{app: snapshots}).
}
The eccentricities and inclinations of planetesimals in the co-rotation resonances ($a \sim a_\mathrm{p}$) remain small.
A small number of planetesimals are trapped in the co-rotation resonances when $e_\mathrm{p}=0.3$, but no planetesimals are trapped around the more eccentric planets $e_\mathrm{p}\geq0.5$.
Due to the planetesimals trapped in the co-rotation resonances, the mean eccentricities and inclinations are smaller when $e_\mathrm{p}=0.1$ than in the other cases. 
The trapped planetesimals gradually escape from the resonance.
By the end of the simulations, most planetesimals escape from the resonance and follow the same orbital evolution as the others.

\subsubsection{Effects on our model}
For the above reasons, our orbital evolution model is valid only when $\left \langle e \right \rangle \lesssim e_\mathrm{p}$.
After reaching $\left \langle e \right \rangle \sim e_\mathrm{p}$, orbital excitation gets slower, which makes the mean ejection time of planetesimals $\tau_\mathrm{eject}$ longer than eq.~(\ref{eq:stay_time}).
Therefore, the accretion probability is larger than that our model expects when $e_\mathrm{p} \lesssim 0.3$.
Our accretion model underestimates the accretion probability by a factor of 3 (2) when $e_\mathrm{p}=0.1$ ($0.3$).
This inconsistency gets smaller as the planetary eccentricity gets larger.

\subsubsection{Perturbations from other planets}
\textbf{
In sec.~\ref{sec: EccentricityEvolution} we neglect the gravitational perturbation from planets other than the innermost planet. 
The strength of the eccentricity perturbation $\Delta e_\mathrm{pass}$ depends on $\propto M_\mathrm{p}/{q_p}^{5/2}$ as shown in eqs.~(\ref{eq:ecc_PassingStar}) and (\ref{eq:ecc_PassingStar2}). 
If two Jupiter-mass planets are separated by 10 Hill radius, the inner planet has $\sim 5$ times stronger perturbations.
If the outer planet is heavier than the inner planet, the perturbation from the outer planet could be comparable to the one from the inner planet.
However, in that case, the outer planet is expected to remain in a low eccentricity orbit because of the mass difference. 
Therefore, only the inner planet enters the highly eccentric orbit and dominates the scattering of the planetesimal disk since $\Delta e_\mathrm{pass}$ strongly depends on $q_p$.
The gravitational perturbation from the outer planets might accelerate the excitation of the planetesimal disk; however, this is expected to be a minor effect. 
}

\textbf{On the other hand, we found that resonances with the innermost planet play a role in planetesimal accretion.
The gravitational perturbation from the other planet might alter the resonance configuration and the accretion probability. 
If the eccentricity excitation of the planet is as slow as $\tau_\mathrm{ecc}=10^5 P_\mathrm{orb,p}$, almost all planetesimals are affected by the secular resonances as shown in Sec.~\ref{sec: EccentricityEvolution}.
If the outer planets completely break secular resonances, the planetesimals would be ejected before the innermost planet crosses their orbit, preventing planetesimal accretion. 
On the other hand, when $\tau_\mathrm{ecc}=10^3 P_\mathrm{orb,p}$, the accretion probability would not change since the secular resonances do not affect the accretion probablity in this case.
}

\textbf{
Also, we do not consider the orbital swap between planets. 
\citet{Mustill+2018} ran N-body simulations with three planets and a planetesimal disk, where the planet initially located in the innermost orbit could be scattered into the outermost orbit. 
Since their main focus was to investigate the accretion onto the white dwarf, it is unclear how many planetesimals accreted onto each planet. 
Future studies should assess these effects by running the N-body simulations with several planets.
}

\subsection{Enrichment of eccentric giant planets}\label{sec: discussion_warm_Jupiter}

\begin{figure}
  \begin{center}
    \includegraphics[width=80mm]{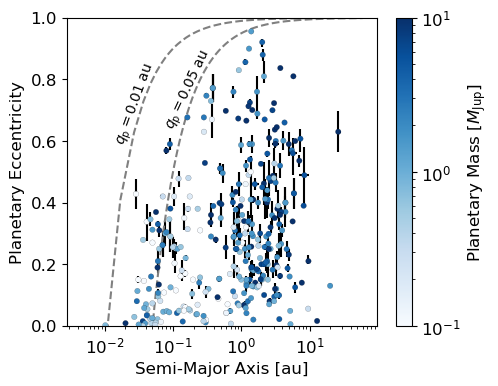}
    \caption{    
    Semi-major axis vs.~ eccentricity of observed exoplanets.
    \textbf{
    The data is obtained from Data from Data \& Analysis Center for Exoplanets (DACE) in April 2024.
    Here, we include exoplanets heavier than Neptune, with a measurement uncertainty of less than 20 \% in eccentricity. 
    }
    The grey dashed lines show the boundary of the regions where planets could be tidally disrupted ($q_\mathrm{p}< 0.01 \au$) and where tidal forces with the central star are sufficient to cause inward planetary migration ($q_\mathrm{p}< 0.05 \au$).   }
    \label{fig: axip_eccp}
  \end{center}
\end{figure}

Giant planets in exoplanetary systems have a wide range of eccentricities (see fig.~\ref{fig: axip_eccp}). 
Data from Data \& Analysis Center for Exoplanets (DACE) show that Jupiter-mass planets (with $M_\mathrm{p}>0.04 M_\mathrm{Jup}$) have an average eccentricity $\simeq0.30$ with a standard deviation of $\simeq0.22$, and a median eccentricity of $\simeq0.26$.
The eccentricity distribution of eccentric Jupiters could be generated by the orbital instability between giant planets after disk dissipation \citep{Carrera+2019, Anderson+2019, Garzon+2022}.
If solid disks exist, these eccentric Jupiters can accrete solids during their dynamical history. 
Below, we estimate the solid accretion during the eccentricity excitation process.

\subsubsection{Accretion of remnant solid disk onto cold Jupiters}\label{sec: discussion_SolidDisk}

We consider a planetary system with multiple giant planets that initiates orbital instability and estimate the mass of accreted solids onto the innermost planet $M_\mathrm{acc}$. 
By the time the planet has reached its current eccentricity $e_\mathrm{p}$, the mass of accreted solids is given by: 
\begin{eqnarray}
    M_\mathrm{acc} = \int_{r_\mathrm{in}}^{r_\mathrm{out}} 2 \pi a_0 \Sigma_\mathrm{sol} P_\mathrm{acc} (a_\mathrm{0}) \mathrm{d} a_\mathrm{0}, \label{eq: Macc_WarmJupiter}
\end{eqnarray}
where $\Sigma_\mathrm{sol}$ is the surface density of solid material. 
For simplicity, we set $\Sigma_\mathrm{sol}$ as a constant. 
We set $r_\mathrm{in}$ to be the planet's pericenter  $a_\mathrm{p} (1-e_\mathrm{p})$ and $r_\mathrm{out}$ to the inner edge of the planetary feeding zone  $a_\mathrm{p}(1-2 \sqrt{3}(M_\mathrm{p}/3M_\mathrm{s})^{1/3})$ because the planet scatters planetesimals inside the feeding zone before the disk dissipates \citep[e.g.][]{Shibata+2019, Shibata+2022c}.
We fix the planet's semi-major axis during the integration of eq.~(\ref{eq: Macc_WarmJupiter}).
However, mutual scattering between the giant planets changes the planetary semi-major axis. 
If the planet is scattered inward, which is a natural outcome of the planet-planet scattering, the planet sweeps planetesimals orbiting interior to  $r_\mathrm{in}$ and accretes more solids. 
Also, the planetesimals orbiting beyond the planet would be scattered by the other giant planets. 
Some scattered planetesimals could accrete to the innermost planet as shown in \citet{Matter+2009}, but we neglect this contribution. 
A comparison between our inferred $M_\mathrm{acc}$ and their results is presented below. 

\begin{figure*}
  \begin{center}
    \includegraphics[width=150mm]{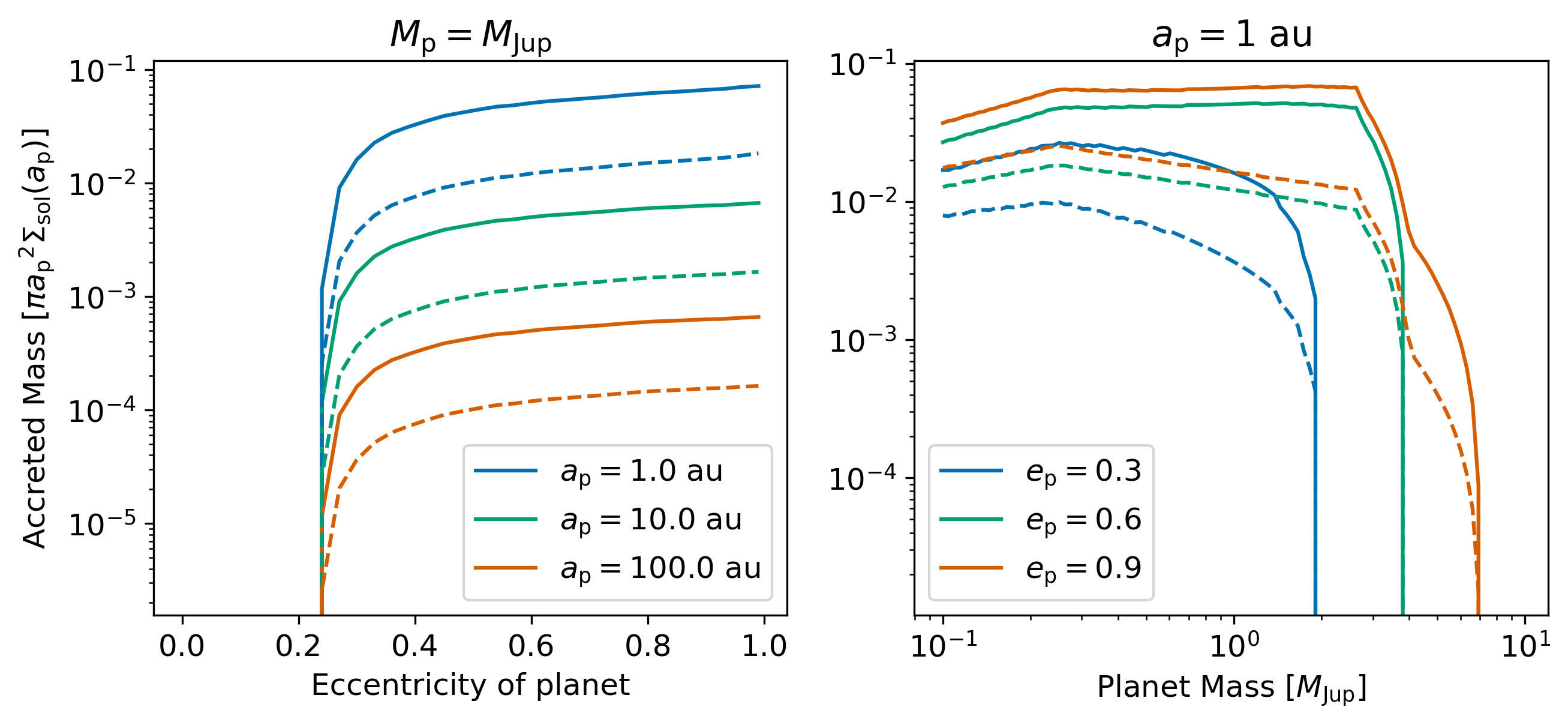}
    \caption{
    Estimated mass of accreted solid materials $M_\mathrm{acc}$ normalised with $\pi {a_\mathrm{p}}^2 \Sigma_\mathrm{sol}$ for various planets.
    The left panel shows $M_\mathrm{acc}/\pi {a_\mathrm{p}}^2 \Sigma_\mathrm{sol}$ as a function of the planetary eccentricty.
    The different color corresponds to the different semi-major axis of the planet.
    The planetary mass is set to Jupiter's mass.
    The right panel shows $M_\mathrm{acc}/\pi {a_\mathrm{p}}^2 \Sigma_\mathrm{sol}$ as a function of the planetary mass.
    Here, the different color corresponds to the different eccentricity of the planet.
    The semi-major axis is set to 1 au.   
    In both panels, the solid and dashed lines show the cases of $\sin i_\mathrm{p} = 10^{-2}$ and $10^{-1}$, respectively.
    We set $\tau_\mathrm{ecc}$ to $10^3 P_\mathrm{orb,p}$ assuming the planet-planet scattering.
    }
    \label{fig: Macc_Multi}
  \end{center}
\end{figure*}

Figure~\ref{fig: Macc_Multi} shows the inferred mass of accreted solids.
We normalise $M_\mathrm{acc}$ with $\pi {a_\mathrm{p}}^2 \Sigma_\mathrm{sol}$.
Here, we adopt the mass-radius (M-R) relation from \citet{Otegi+2020} and use Jupiter's radius as an upper limit of the planetary radius.
The planetary radius is set as: 
\begin{eqnarray}
    R_\mathrm{p} = \min \left( 0.70 R_\oplus \left( \frac{M_\mathrm{p}}{M_\oplus} \right)^{0.63} , R_\mathrm{Jup} \right).
\end{eqnarray} 
The solid and dashed lines show the cases with a planetary inclination of $\sin i_\mathrm{p}=10^{-2}$ and $10^{-1}$, respectively.
We set the eccentricity evolution timescale to  $\tau_\mathrm{ecc}=10^3 P_\mathrm{orb,p}$ assuming planet-planet scattering.
The left panel shows $M_\mathrm{acc}$ as a function of the planetary eccentricity.
When the planetary eccentricity is smaller than $\sim 0.2$, the eccentric planet cannot accrete planetesimals because the planet's pericenter is within the planetary feeding zone. 
When $e_\mathrm{p} \gtrsim 0.2$, $M_\mathrm{acc}$ increases with planetary eccentricity.
Solid accretion is more effective for small radial distances and lower inclination.
The right panel shows $M_\mathrm{acc}$ as a function of planetary mass.
The mass of accreted solid rapidly decreases when $M_\mathrm{p}>M_\mathrm{Jup}$.
This is because $e_\mathrm{cross}$ is larger for more massive planets.
When $M_\mathrm{p}>2M_\mathrm{Jup}$, $e_\mathrm{cross}$ exceeds $0.3$ and $P_\mathrm{acc}$ decreases.
This means the solids are ejected from the planetary system before the planetary orbit crosses the solid disk.

While it is still unclear how many solids remain around giant planets after disk dissipation, N-body simulations show that many planetesimals (more than 10 $M_{\oplus}$) could remain around a giant planet \citep[e.g.][]{Zhou+2007, Shibata+2019, Shibata+2022c}.
Also, the giant planet itself can generate a new planetesimal disk via a gap opening in the protoplanetary disk \citep{Shibaike+2020, Shibaike+2023, Eriksson+2022}.
After disk dissipation, the remnant planetesimal disk excites itself through the mutual gravitational interactions and initiates collisional cascades \citep{Wyatt+2007, Lohne+2008}.
\citet{Heng+2010} found that the maximum surface density of solids around an old star of $3$ Gyrs exceeds $\Sigma_\mathrm{sol} = 1 \g/\cm^2$ beyond $10 \au$.
Also, the observed dust emission suggests the existence of a massive solid disk at distances of  $10-100 \au$ \citep{Krivov+2020}. 
If a solid disk of $\Sigma_\mathrm{sol} = 1 \g/\cm^2$ remains at $10 \au$, $\pi {a_\mathrm{p}}^2 \Sigma_\mathrm{sol}$ is $12 M_\oplus$ and $M_\mathrm{acc}$ is $0.01-0.1 M_\oplus$.
This heavy-element mass can't increase the planetary bulk metallicity but could increase the atmospheric metallicity (\citet{Howard+2023, Muller+2024}). 
Therefore, we predict that the atmosphere of eccentric cold Jupiters will be enriched with heavy elements, but the actual enrichment depends on the planetary eccentricity and mass.
Solid accretion does not occur for massive cold Jupiter of $M_\mathrm{p}>7 M_\mathrm{Jup}$.

In this study, we use a single eccentric planet, although we assume that the planet is excited by other giant planets. 
The existence of the other giant planets would increase the accretion of solids because they scatter planetesimals from the outer orbit inwards. 
\citet{Matter+2009} showed that the accretion of comets scattered by Uranus and Neptune brings $\sim 0.1 M_\oplus$ of heavy elements to Jupiter.
Since Jupiter has a low eccentricity $<0.1$ in their model, comet accretion onto an eccentric cold Jupiter would be smaller than $0.1 M_\oplus$.
Nevertheless, the scattering of the remnant solid disk by multiple giant planets would increase the mass of the accreted solid after disk dissipation.
This process should be addressed in future studies.

The available solid surface density would be smaller than $\Sigma_\mathrm{sol} = 10^{-3} \g/\cm^2$ at $1 \au$ \citep{Heng+2010} because the collisional evolution timescale strongly depends on the radial distance from the central star. 
Despite the higher accretion efficiency, the mass of accreted solids would be smaller than $10^{-5} M_\oplus$ in the inner orbit. 
Therefore, other solid accretion processes like comet accretion \citep{Seligman+2022} would be required to enrich the planet further.

\subsubsection{Collisions of planets}\label{sec: discussion_Embryos}

\begin{figure*}
  \begin{center}
    \includegraphics[width=150mm]{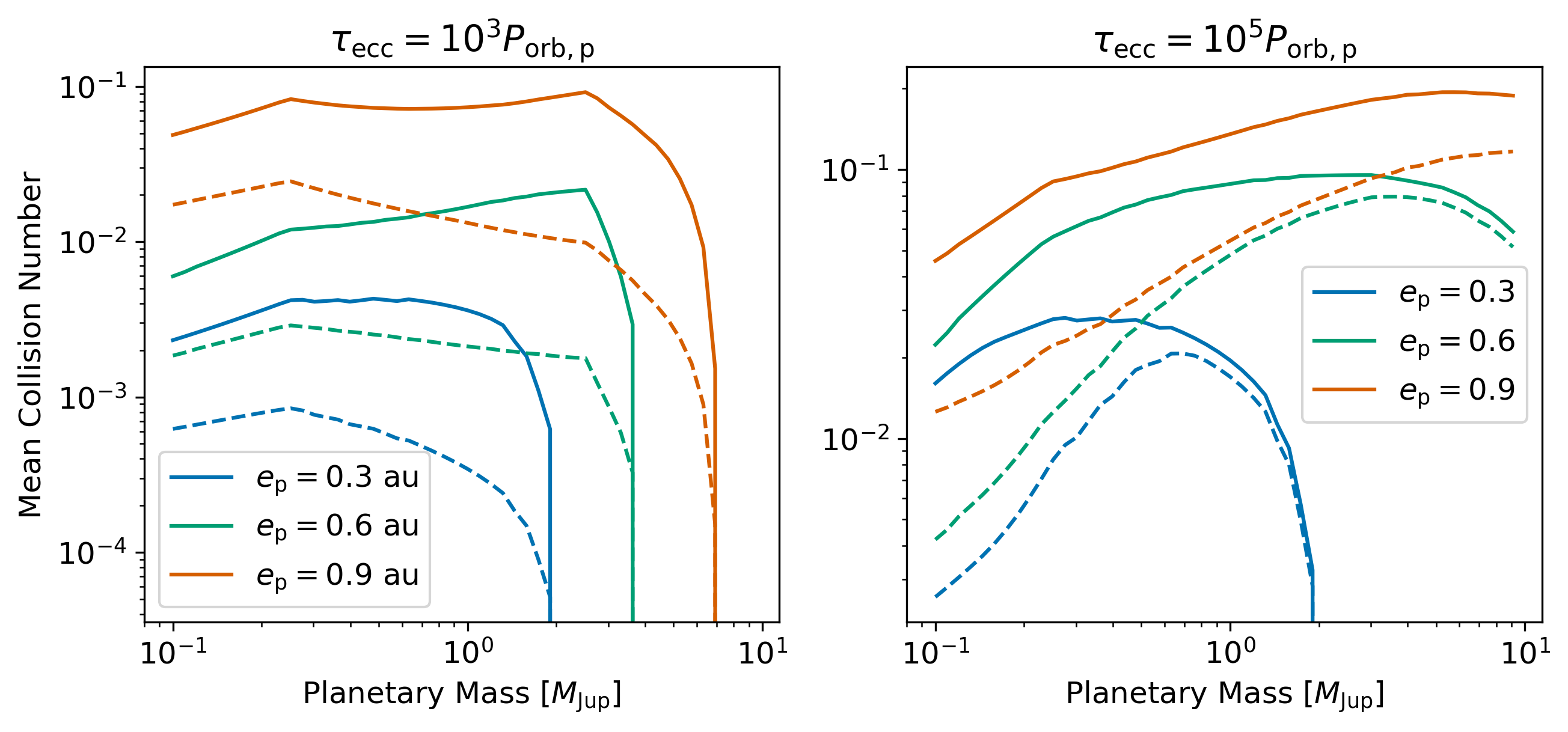}
    \caption{
    Estimated mean collision number of planets onto an eccentric planet.
    We show the mean collision number as a function of the planetary mass.
    The left and right panel show the cases with $\tau_\mathrm{ecc}=10^3 P_\mathrm{orb,p}$ and $\tau_\mathrm{ecc}=10^5 P_\mathrm{orb,p}$, respectively.
    The blue, green, and orange lines correspond to $e_\mathrm{p}=0.3$, $0.6$, and $0.9$, respectively.
    The solid and dashed lines show the cases of $\sin i_\mathrm{p} = 10^{-2}$ and $10^{-1}$, respectively.
    }
    \label{fig: Memb}
  \end{center}
\end{figure*}

The statistical analysis shows that many planetary systems with a cold Jupiter have inner terrestrial planets \citep{Zhu+2018, Bryan+2019, Rosenthal+2022}.
Collisions of such inner terrestrial planets could increase the planetary metallicity of eccentric giant planets.
\citet{Mustill+2015} used N-body simulations to show that some inner planets collide with the outer giant planets during orbital instability. 
In this section, we estimate the collision probability between eccentric Jupiters and inner terrestrial planets.  
\citet{Kunimoto+2020} derive a fitting function for the occurrence rate of planets with $R_\mathrm{p}<4 R_\oplus$ in the inner orbit, which is given by: 
\begin{eqnarray}\label{eq: Occurence_Rate}
    F_\mathrm{occ} &=& \frac{\mathrm{d} f}{\mathrm{d} \log P_\mathrm{orb,p}} \nonumber \\
    &=& C_\mathrm{KM20}{P_\mathrm{orb,p}}^{\beta_\mathrm{KM20}}  \left( 1 - \gamma_\mathrm{KM20} \exp \left(-\frac{P_\mathrm{orb,p}}{P_\mathrm{orb,0}}\right) \right),
\end{eqnarray}
where $C_\mathrm{KM20}$, $\beta_\mathrm{KM20}$, $\gamma_\mathrm{KM20}$, and $P_\mathrm{orb,0}$ are fitting parameters (see \citet{Kunimoto+2020} for each value).
 
Note that eq.~(\ref{eq: Occurence_Rate}) gives the occurrence rate for all planetary systems, but a planetary system with a cold Jupiter would have a larger occurrence rate for the inner planets \citep{Zhu+2018, Bryan+2019, Rosenthal+2022}. 
The mean collision number of inner planets $N_\mathrm{col}$ can be obtained by integrating $\int_{r_\mathrm{in}}^{r_\mathrm{out}} F_\mathrm{occ} P_\mathrm{acc} \mathrm{d} a_0$. 
Figure~\ref{fig: Memb} shows the estimated mean collision number for the inner planets.
Here, we consider the case of $a_\mathrm{p}=1 \au$ and use the same M-R relation as sec.~\ref{sec: discussion_SolidDisk}.
The results are presented for different assumed eccentricity excitation timescales of  $\tau_\mathrm{ecc}=10^3 P_\mathrm{orb,p}$ (left panel) and $\tau_\mathrm{ecc}=10^5 P_\mathrm{orb,p}$ (right panel), and different planetary inclination $\sin i_\mathrm{p}=0.01$ (solid lines) and $\sin i_\mathrm{p}=0.1$ (dashed lines).
We find that $N_\mathrm{col}$increases with increasing  $e_\mathrm{p}$ and $\tau_\mathrm{ecc}$, and for decreasing $\sin i_\mathrm{p}$.

With the averaged eccentricities of detected Jupiter-mass planets $\simeq0.3$, $N_\mathrm{col}$ is estimated to be smaller than 0.04.  
Therefore, collisions of inner terrestrial planets onto the eccentric Jupiter would be minor events.
On the other hand, hot Jupiters might experience collisions with a higher probability.
One of the possible formation pathways of hot Jupiters is high-eccentric migration.
A giant planet perturbed to a high-eccentric orbit would dissipate its orbital energy and migrate inward through interaction with the central star. 
A hot Jupiter formed by high-eccentric migration should have been excited to the highly eccentric orbit of $e_\mathrm{p}\gtrsim 0.9$ if the initial semi-major axis was $\sim 1 \au$.
Using N-body simulations \citet{Garzon+2022} found that most highly eccentric planets are formed via slow eccentricity excitation mechanisms like secular interactions.
Assuming that hot Jupiters are formed via high-eccentric migration with $e_\mathrm{p}\gtrsim0.9$ and $\tau_\mathrm{ecc}\gtrsim10^5 P_\mathrm{orb,p}$, we predict that roughly 10 \% of the hot Jupiters collided with inner terrestrial planets. 

The dataset used in \citet{Kunimoto+2020} shows a peak at $R_\mathrm{p}\sim2.5 R_\oplus$ in the occurrence rate. 
This corresponds to a mass of $M_\mathrm{p}\sim10M_\oplus$ from the M-R relation of  \citet{Otegi+2020}. 
A collision of a $10M_\oplus$ planet could increase the planetary core mass and bulk metallicity.
However, such collision probability is only of the order of $\sim10\%$ and therefore cannot explain the inferred high metallicity of hot Jupiters \citep[e.g.][]{Guillot+2006}.  
Other heavy element accretion processes such as planetesimal accretion \citep{Hands+2021, Knierim+2022} and accretion of enriched gas \citep{Schneider+2021, Schneider+2021b} before disk dissipation are required to explain the enrichment of hot Jupiters. 

\subsection{Pollution of the central star}
\begin{figure}
  \begin{center}
    \includegraphics[width=80mm]{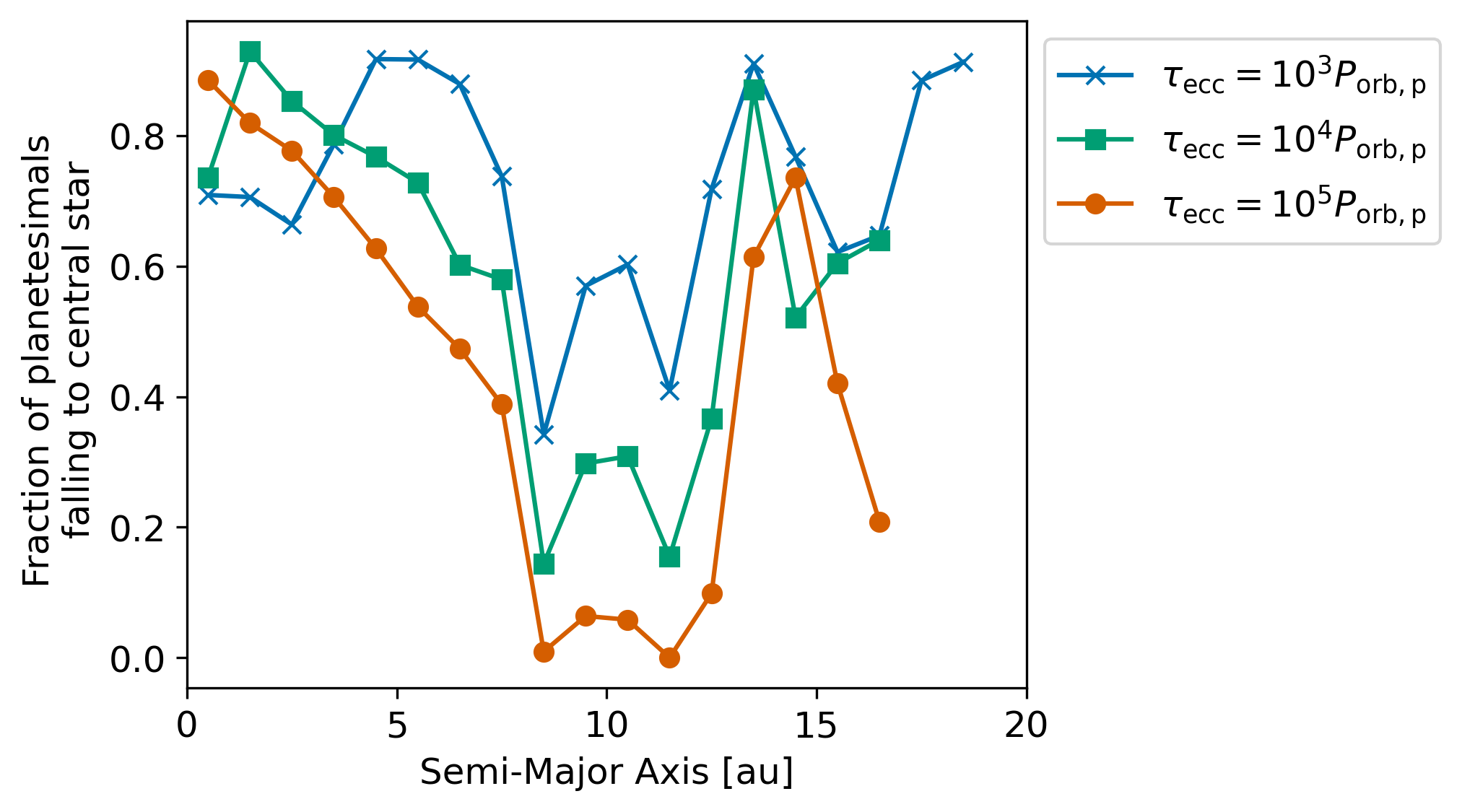}
    \caption{
    Fraction of planetesimals that fall onto the central star as a function of the planetesimals' initial semi-major axis.
    The blue, green, and orange lines show the cases of $\tau_\mathrm{ecc}=10^3 P_\mathrm{orb,p}$, $10^4 P_\mathrm{orb,p}$, and $10^5 P_\mathrm{orb,p}$, respectively.
    }
    \label{fig: Ffall}
  \end{center}
\end{figure}

Our simulations show that most of the remaining solids fell onto the central star after disk dissipation.
Figure~\ref{fig: Ffall} presents the fraction of planetesimals that fall onto the star $F_\mathrm{fall}$ during the planetary eccentricity evolution (using the results obtained in sec.~\ref{sec: EccentricityEvolution}).
\textbf{
\citet{Rodet+2023} find that $F_\mathrm{fall}$ is larger for a planet with higher eccentricity.
In our simulations, the planet is found to scatter planetesimals between the pericenter and apocenter, and this region expands with the planetary eccentricity.
Therefore, $F_\mathrm{fall}$ is small around the planet's semi-major axis and increases as the orbital separation from the planet increases.
\citet{Rodet+2023} fixed the planet's eccentricity, the same as the simulations in Sec.~\ref{sec: CollisionToPlanet_SIM}.
Comparing Fig.~\ref{fig:fate_standard} and Fig.~\ref{fig: Ffall}, we find that $F_\mathrm{fall}$ decreases when we include the effect of eccentricity evolution and with increasing eccentricity excitation timescale $\tau_\mathrm{ecc}$.
Our results imply that the pollution of the central star depends on the orbital evolution history of their giant planets. We hope to investigate this in further detail in future research. 
}

The observed enhancement of Li in Sun-like stars implies that a significant amount of solids polluted the stellar atmosphere during their lifetime \citep[][see references there]{Spina2024}. 
While the engulfment of planets is suggested as a source of stellar atmospheric pollution \citep{Spina+2021}, 
accretion of scattered planetesimals onto the central star is another possible mechanism for enriching stellar atmospheres with refractory materials.  
Planetesimal accretion onto a central star has been investigated by various authors using N-body simulations (\citet{Beust+2000, Frewen+2014, Mustill+2018, Smallwood+2021, Veras+2022, OConnor+2022, Rodet+2023}, see also \citet{Veras2021} for more references). 
\citet{Mustill+2018} investigate the accretion of the remnant planetesimal disk onto white dwarfs and find that $13.3 \%$ of planetesimals distributed between $5 \au$ and $8.5 \au$ entered the Roche limit of the white dwarf and got accreted. 
Our simulations suggest a higher accretion probability of $64 \%$ for $\tau_\mathrm{ecc}=10^3 P_\mathrm{orb,p}$ and $38 \%$ for $\tau_\mathrm{ecc}=10^5 P_\mathrm{orb,p}$ in the same orbital region. 
\textbf{
As shown in \citet{Rodet+2023}, the accretion probability is higher for a larger stellar radius and a larger stellar mass.
In this study, we use a larger physical radius of the central star than the radius of the Roche limit of the white dwarf and a larger stellar mass. As a result, the accretion probability we infer is larger in comparison to \citet{Mustill+2018}.
}
This implies that the pollution of the stellar atmosphere is more probable for solar-type stars than white dwarfs.
\citet{Spina2024} shows that the accretion of several $M_\oplus$ of solids onto a solar-type star increases the iron fraction in the outer convective cell beyond the detection limit and that this enrichment can be observed because the enriched layer persists for $\sim 1$ Gyrs  \citep{Behmard+2022}.
Our study also suggests that several $M_\oplus$ of planetesimals could be accreted by the star after the onset of orbital instability of giant planets if a solid disk of $\Sigma_\mathrm{sol}=1 \g/\cm^2$ remains around 10 $\au$. 
However, since our orbital evolution model is not valid when  $\left \langle e \right \rangle>e_\mathrm{p}$ (see sec.~\ref{sec: discussion_model}), we can't derive an analytical prescription for the accretion process onto the central star. 
As shown in Fig.~\ref{fig: EccJupiter_Resonances}, the orbital evolution of planetesimals, including the effects of secular perturbations, needs to be modeled, and we hope to address this topic in our future research.

\section{Summary and Conclusions}
\label{sec: Summary}
We investigate the possibility of solid accretion onto an eccentric gas-giant planet after disk dissipation.
We focus on the scenario where a giant planet is excited by other giant planets and investigate the accretion probability of solids onto the excited planet.
We find that the orbital evolution of solid particles, such as planetesimals and embryos, is regulated by weak encounters with the eccentric planet rather than strong close encounters. 
We show that even in the region where the Safronov number is smaller than unity, most of the solids fall onto the central star or are ejected from the planetary system.
\par 

We also derive an analytical formula for the accretion probability of solid particles.
Our analytical prescription can be applied for both planetesimals and embryos by setting the initial inclination appropriately.
We also investigate the solid accretion rate of a giant planet whose eccentricity increases with time.
In that case, we find that resonance effects play an important role in the orbital evolution of the solid particles before the planetary orbit crosses the solid disk.
We derive the condition determining when the solid particles are ejected from the planetary system before the orbits are crossed.  
We find that solid accretion mainly occurs in the two orbital regions:
the inner disk where the planetary capture radius is larger than the disk's thickness, and at the planet's vicinity where the planet's orbit crosses that of planetesimals when gravitational focusing is efficient.

The main conclusions from our study can be summarized as follows:
\begin{itemize}
    \item[1.] Eccentric Jupiters around $10 \au$ could accrete $0.01-0.1 M_\oplus$ of heavy elements from the solid disk remaining around the planet. 
The accretion probability depends on the planetary eccentricity, inclination, and mass. Therefore the metallicity of the planetary atmosphere should relate to these parameters. 
While the accretion probability is higher in the inner orbit, the accretion of the solid disk is negligible around $1 \au$ because the available solid disk mass is too small in the inner orbit. \item[2.] Using the occurrence rate of terrestrial planets, we estimate how many terrestrial planets collide with an  eccentric planet.
The number of planets colliding with the eccentric planet increases with the planetary eccentricity.
If hot Jupiters formed via high-eccentric migration, $\sim 10 \%$ of hot Jupiters were impacted by a planet of $\sim 10 M_\oplus$.
\item[3.] We find that several Earth masses of solid material scattered by the eccentric planet could accrete onto the central star.
Accretion of the scattered solid disk would be a source of heavy elements of the stellar atmosphere as well as the engulfment of planets.
\end{itemize}

We conclude that solid accretion onto an eccentric planet would be a minor process for planets in the inner orbit $\lesssim 1 \au$.
Therefore, a different solid accretion process, such as the accretion of small bodies such as comets and asteroids would be required to enrich the planetary atmosphere with heavy elements.
This work represents the first step in investigating how the orbits of planetesimals and embryos evolve around a single eccentric planet. However, since some planetary systems can have more than one giant planet, future studies should explore solid accretion accounting for multiple giant planets. 
Our study clearly shows that in order to understand the observed composition of giant planets, and even stars, a robust understanding of the early evolution of planetary systems is required.

\section*{Acknowledgements}
{
We thank the anonymous reviewer for constructive comments. 
We thank Hiroshi Kobayashi for the fruitful discussion about the dynamics part of this study.
SS and RH acknowledge support from the Swiss National Science Foundation (SNSF) under grant $200020\_188460$.
Numerical computations were carried out on the Cray XC50 at the Center for Computational Astrophysics, National Astronomical Observatory of Japan.
}

\appendix

\section{Benchmarks of IAS15}
\label{app: Integrator}
\begin{figure*}
  \begin{center}
    \includegraphics[width=160mm]{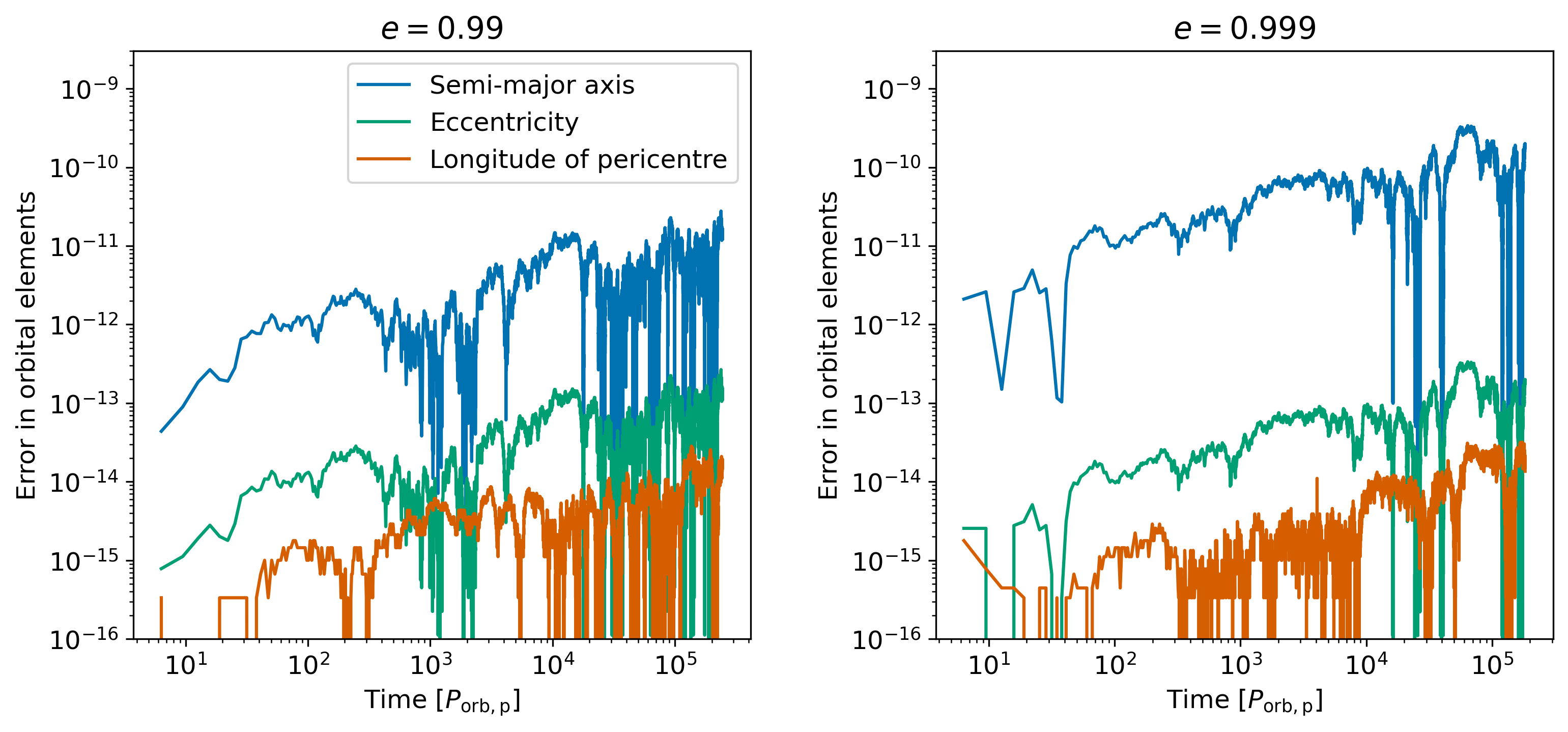}
    \caption{
    Time evolution in the errors of orbital elements.
    Blue, green, and orange lines show the error in the semi-major axis, eccentricity, and longitude of the pericentre, respectively.
    }
    \label{fig: Time_errors}
  \end{center}
\end{figure*}

We investigate the orbital evolution of planetesimals and embryos around an eccentric planet.
The orbital integration of eccentric objects should be carefully treated because the angular velocity changes orders of magnitude in one orbit.
We adopt the IAS15 integrator \citep{Everhart+1985, Rein+2015}, which uses the adaptive timestep.
Figure~\ref{fig: Time_errors} shows the time evolution of numerical errors of an eccentric particle.
Here, we solve the two-body problems of a solar-mass star and an eccentric particle with a semi-major axis of 10 au and eccentricities of 0.99 (left) and 0.999 (right), respectively.
The numerical error is less than $10^{-9}$ even if we integrate them for $10^5$ Kepler orbits.

\section{Resonances of the eccentric planet}
\label{app: resonances}
\begin{figure*}
  \begin{center}
    \includegraphics[width=160mm]{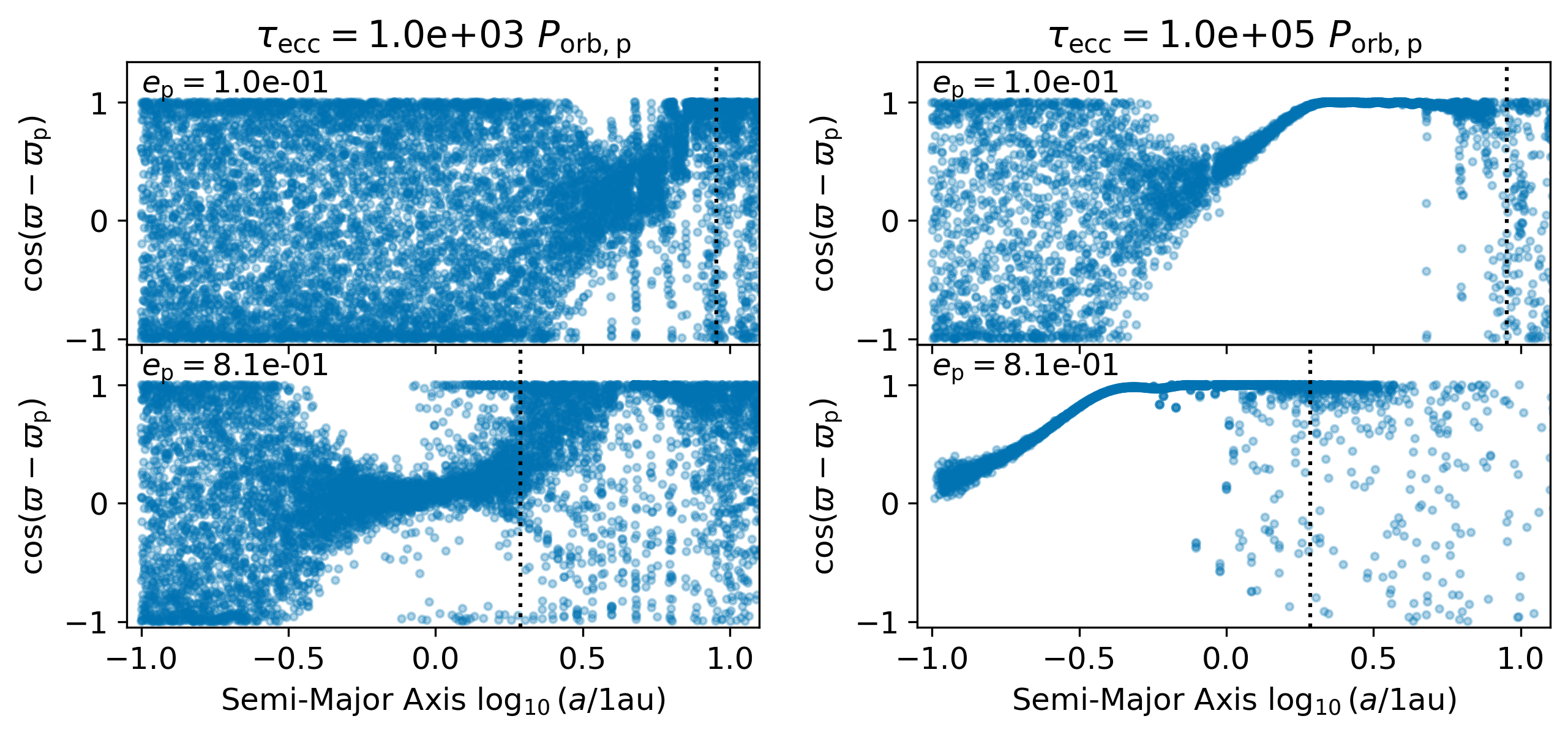}
    \caption{
    Same as Fig.~\ref{fig: Snapshots_psdEccp}, but shows the difference in the pericenters between planetesimals and the planet instead of planetesimals' eccentricity.
    The left and right columns show the cases with $\tau_\mathrm{ecc}=10^3 P_\mathrm{orb,p}$ and $10^5 P_\mathrm{orb,p}$, respectively.
    The upper and lower panels show the snapshots when the planetary eccentricity is $e_\mathrm{p}=0.1$ and $0.81$, respectively.
    The vertical dotted lines are the pericenter and apocenter of the planet.
    }
    \label{fig: Snapshots_Resonance}
  \end{center}
\end{figure*}

\begin{figure*}
  \begin{center}
    \includegraphics[width=160mm]{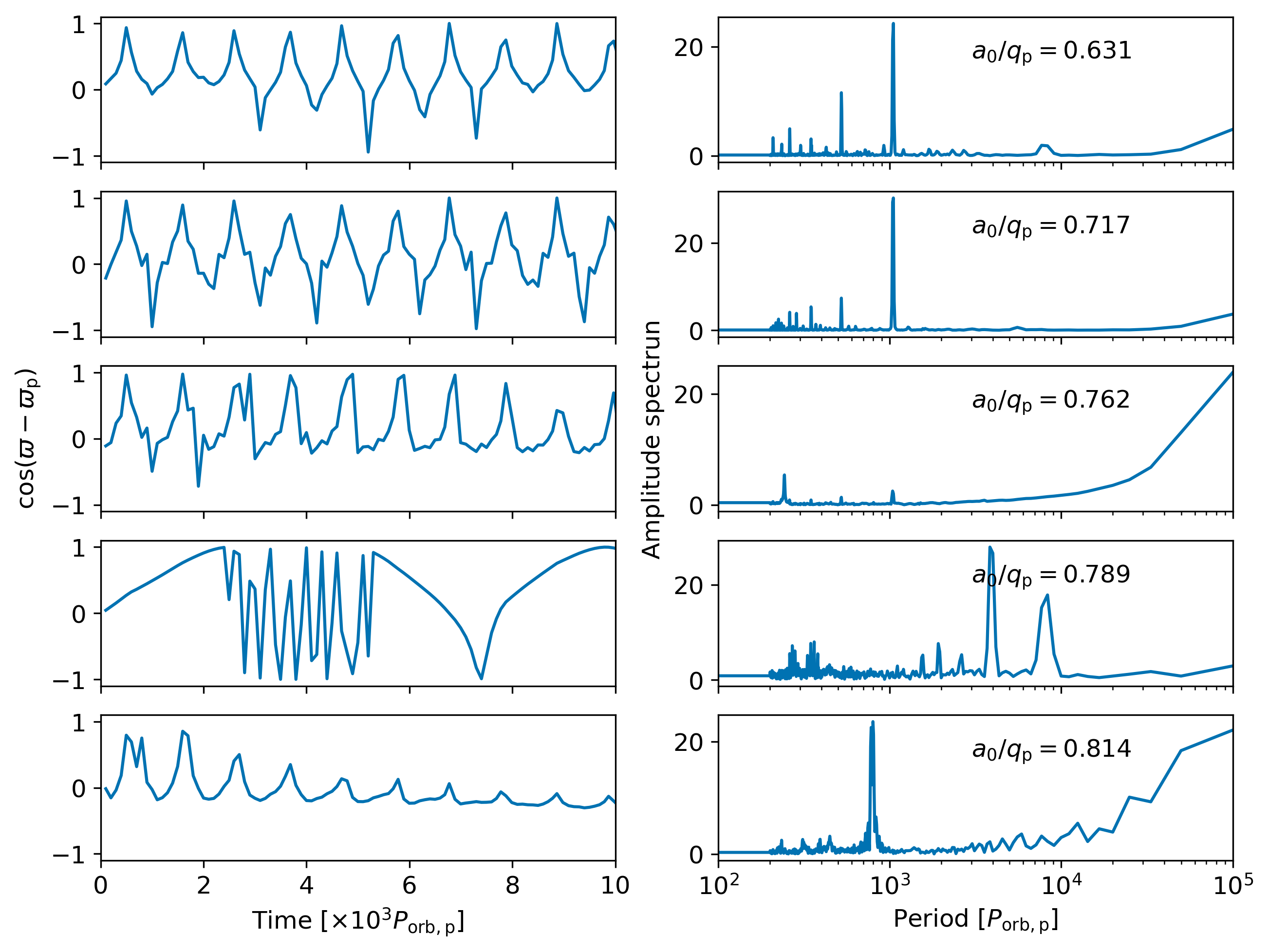}
    \caption{
    {\bf Left columns}: Time evolution of $\sin (\varpi - \varpi_\mathrm{p})$ for each planetesimal.
    {\bf Right columns}: Amplitude spectrum obtained with Fast Fourier Transform.
    The amplitude spectrum is scaled with $N/2$, where N is the number of data points.
    Each planetesimal has a different initial semi-major axis, and it is shown in the upper right corner of the panels in the right column divided by the pericenter of the planet $q_\mathrm{p}=5 \au$.
    }
    \label{fig: FFTs}
  \end{center}
\end{figure*}

As shown in Sec.~\ref{sec: OrbitalEvolutionBOC}, resonances of the eccentric planet play important roles in the orbital evolution of planetesimals.
Figure~\ref{fig: Snapshots_Resonance} shows the difference in the longitude of pericenters between planetesimals and the planet $\varpi - \varpi_\mathrm{p}$.
$\cos (\varpi - \varpi_\mathrm{p}) = 1$ means that the pericenter of the planetesimal and the planet is aligned and the gravitational scattering is weaker than in the case where the pericenters are not aligned.
Therefore, the eccentricity excitation rate gets slower than the eq.~(\ref{eq:ecc_PassingStarEvolv}) expects.
The alignment of the pericenter around a giant planet within a gaseous disk is shown by \citet{Guo+2021, Guo+2022}.
The alignment also occurs around a giant planet whose eccentricity is increasing.

In order to investigate the typical timescale of the resonances, we analyze the orbital evolution of planetesimals around an eccentric Jupiter-mass planet with $a_\mathrm{p}=10 \au$ and $e_\mathrm{p}=0.5$.
Figure~\ref{fig: FFTs} shows the time evolution of $\cos (\varpi - \varpi_\mathrm{p})$ and the amplitude spectrum obtained with the Fast Fourier Transform.
In the amplitude spectrum, some planetesimals have a strong peak around $\sim 10^3 P_\mathrm{orb,p}$.
Therefore, the secular perturbations from the eccentric planet align the planetesimal's pericenter within a timescale of $\sim 10^3 P_\mathrm{orb,p}$.

\section{Snapshots of planetesimals}
\label{app: snapshots}

\begin{figure*}
  \begin{center}
    \includegraphics[width=180mm]{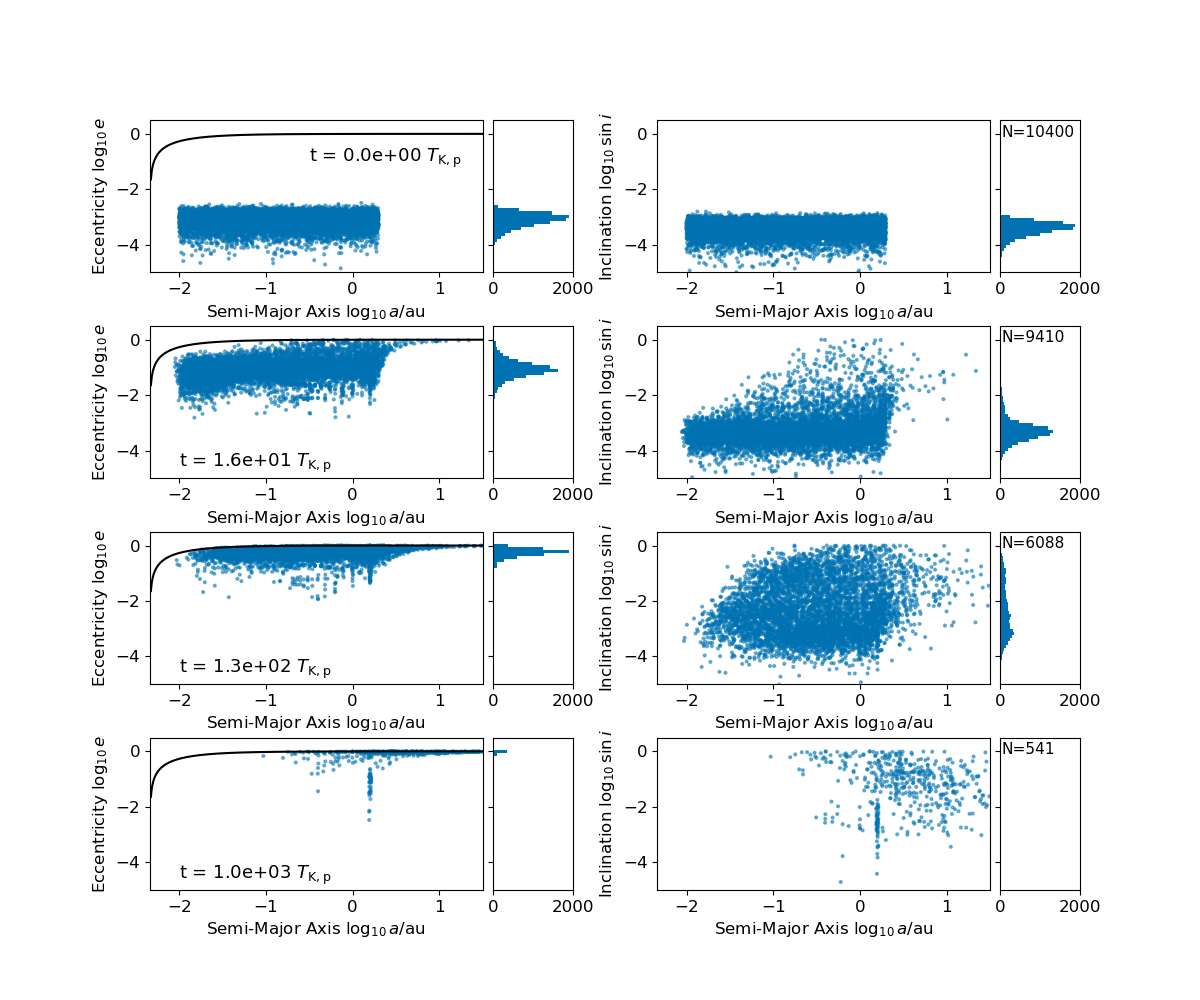}
    \caption{
    Snapshots of orbital integration around an eccentric giant planet.
    The left and right columns show the eccentricity and inclination, respectively, with the semi-major axis of planetesimals.
    Time progresses from upper to lower panels.
    The solid lines in the left column show $e_\mathrm{fall}$.
    The small boxes on the right side of each panel show the histogram.
    The total number of planetesimals $N$ is written in the boxes.
    Here, we show the case with $a_\mathrm{p}=1 \au$, $e_\mathrm{p}=0.99$, $\sin i_\mathrm{p}=0$, $M_\mathrm{p}=M_\mathrm{Jup}$ and $R_\mathrm{p}=R_\mathrm{Jup}$.
    }
    \label{fig: snapshots}
  \end{center}
\end{figure*}

\textbf{
We show the snapshots of our simulations as supporting information for this paper.
Figure~\ref{fig: snapshots} shows the snapshots from our simulation.
We present the case with a planet of $M_\mathrm{p}=M_\mathrm{Jup}$, $R_\mathrm{p}=R_\mathrm{Jup}$, and an orbit with $a_\mathrm{p}=1 \au$, $e_\mathrm{p}=0.99$, $\sin i_\mathrm{p}=0$.
The eccentricity of the entire planetesimal disk is excited by the first passage of the eccentric planet.
As expected in a-particle-in-a-box approximation, the eccentricity excitation rate is independent of planetesimals' semi-major axis.
Eccentricity excitation occurs in every planet's orbit and uniformly within the planetesimal disk.
}

\begin{figure*}
  \begin{center}
    \includegraphics[width=180mm]{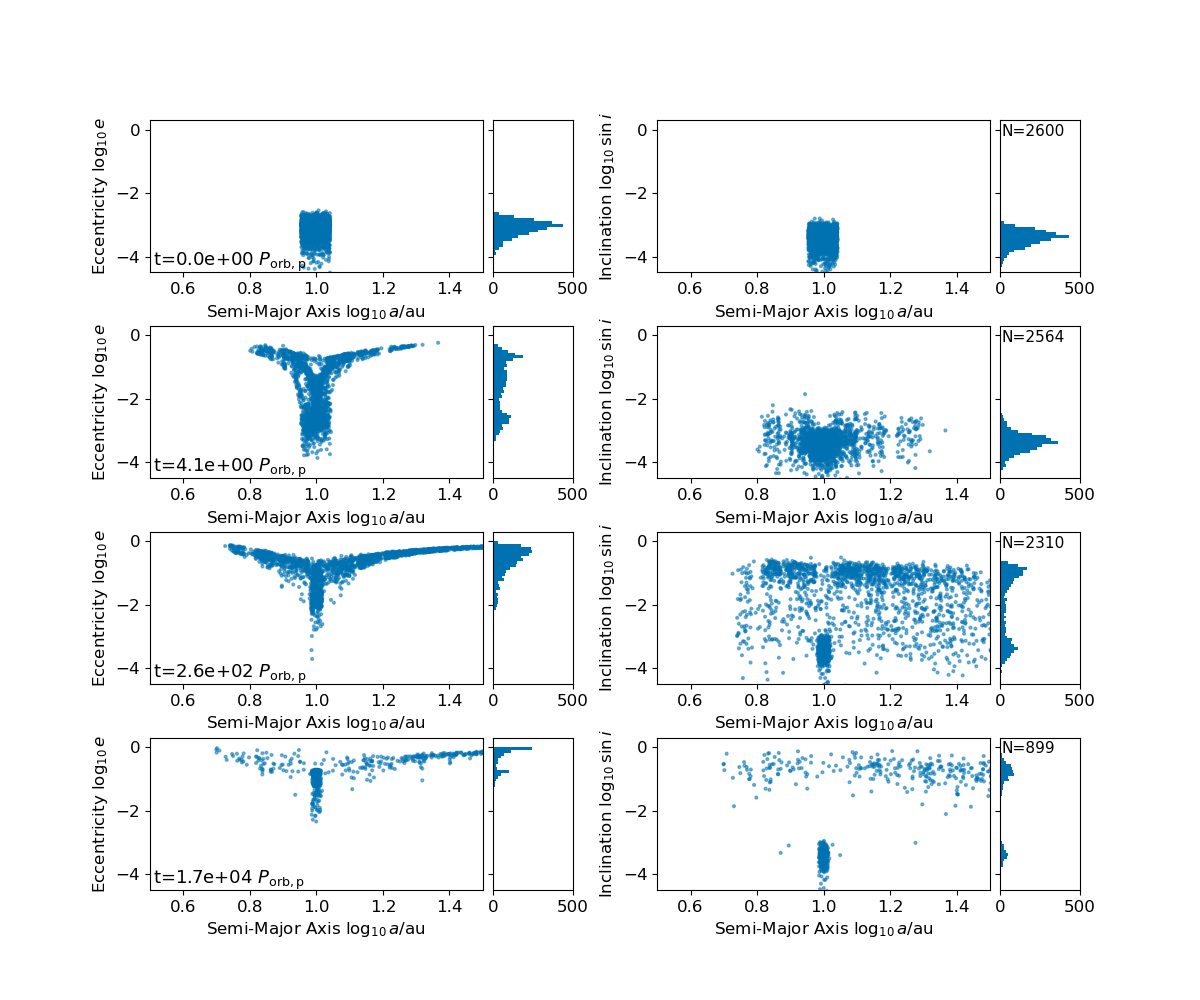}
    \caption{
    Same as Fig.~\ref{fig: snapshots}, but for $e_\mathrm{p}=0.1$.
    Unlike in Fig.~\ref{fig: snapshots}, here, many planetesimals are trapped in the co-rotation resonance of the planet.
    }
    \label{fig: snapshots_ep01}
  \end{center}
\end{figure*}

\textbf{
Figure~\ref{fig: snapshots_ep01} also shows the snapshots from our simulation but with a different eccentricity of the planet $e_\mathrm{p}=0.1$.
In this case, many planetesimals are trapped in the co-orbital resonance around the planet's semi-major axis $a_\mathrm{p}=10\mathrm{au}$.
The eccentricity and inclination excitation is slower in the co-orbital resonance.
Therefore, the mean eccentricity and inclination obtained in the N-body simulations are smaller than the analytical models given by eq.~(\ref{eq:TimeEvolution_Ecc}) and eq.~(\ref{eq:TimeEvolution_Inc}) predict.
}


%
%

\bibliographystyle{aa} %
\bibliography{refs} %

\end{document}